\documentclass{article}
\usepackage[utf8]{inputenc}
\usepackage{graphicx}

\usepackage{amssymb}
\usepackage{amsmath}
\usepackage{amssymb}
\usepackage{lineno}

\usepackage{array}
\newcolumntype{P}[1]{>{\centering\arraybackslash}p{#1}}

\usepackage{graphics}
\usepackage{caption}
\usepackage{subcaption}

\newcommand{\halb}{\frac{1}{2}}
\renewcommand{\d}{{\rm d}}			
\renewcommand{\Re}{\mathsf{Re}}	    
\newcommand{\CFL}{\mathsf{CFL}}		
\renewcommand{\P}{\mathsf{P}}	    
\usepackage{color}

\usepackage{authblk}
\usepackage[margin=3.0cm]{geometry}

\title{Numerical methods for hydraulic transients in visco-elastic pipes}

\author[$\dagger$]{Giulia Bertaglia \footnote{Corresponding author. Email address: \textit{giulia.bertaglia@unife.it}}}
\author[$\star$]{Matteo Ioriatti}
\author[$\dagger$]{Alessandro Valiani}
\author[$\star$]{Michael Dumbser}
\author[$\dagger$]{Valerio Caleffi}

\affil[$\dagger$]{\small Department of Engineering, University of Ferrara, Via G. Saragat 1, 44122 Ferrara, Italy}
\affil[$\star$]{\small Department of Civil, Environmental and Mechanical Engineering, University of Trento, Via Mesiano 77, 38123 Trento, Italy}

\begin{document}
 \maketitle

\begin{abstract}
In technical applications involving transient fluid flows in pipes the convective terms of the corresponding governing equations are generally negligible. Typically, under this condition, these governing equations are efficiently discretised by the Method of Characteristics (MOC). Only in the last years the availability of very efficient and robust numerical schemes for the complete system of equations, such as recent Finite Volume Methods (FVM), has encouraged the simulation of transient fluid flows with numerical schemes different from the MOC, allowing a better representation of the physics of the phenomena.\\
In this work, a wide and critical comparison of the capability of Method of Characteristics, Explicit Path-Conservative Finite Volume Method (DOT solver) and Semi-Implicit (SI) Staggered Finite Volume Method is presented and discussed, in terms of accuracy and efficiency. To perform the analysis in a framework that properly represents real-world engineering applications, the visco-elastic behaviour of the pipe wall, the effects of the unsteadiness of the flow on the friction losses, cavitation and cross-sectional changes are taken into account. \\
The analyses are performed comparing numerical solutions obtained using the three models against experimental data and analytical solutions. In particular, water hammer studies in high density polyethylene (HDPE) pipes, for which laboratory data have been provided, are used as test cases. Considering the visco-elastic mechanical behaviour of plastic materials, a 3-parameter and a multi-parameter linear visco-elastic rheological models are adopted and implemented in each numerical scheme. Original extensions of existing techniques for the numerical treatment of such visco-elastic models are introduced in this work for the first time. After a focused calibration of the visco-elastic parameters, the different performance of the numerical models is investigated. A comparison of the results is presented taking into account the unsteady wall-shear stress, with a new approach proposed for turbulent flows, or simply considering a quasi-steady friction model. A predominance of the damping effect due to visco-elasticity with respect to the damping effect related to the unsteady friction is confirmed in these contexts. Moreover, all the numerical methods show a good agreement with the experimental data and a high efficiency of the MOC in standard configuration is observed.\\
Finally, three Riemann Problems (RP) are chosen and run to stress the numerical methods, taking into account cross-sectional changes, more flexible materials and cavitation cases. In these demanding scenarios, the weak spots of the Method of Characteristics are depicted. 
\end{abstract}



\section{Introduction}
\label{S:1}
Flexible plastic pipes in polyvinyl chloride (PVC), polyethylene (PE) and in particular high density polyethylene (HDPE) are gaining an increasingly important role in pressurized and not pressurized hydraulic systems, being often preferred to other materials (i.e. steel and concrete) for water distribution networks, irrigation plants and sewage systems. This trend is a consequence of the excellent mechanical and chemical properties of polymer materials, even more considering the easy and rapid process of installation required and the cheaper prices. Almost without exception, polymers belong to a class of substances that show visco-elastic properties, responding to external forces in an intermediate manner between the behavior of an elastic solid and a viscous liquid \cite{shaw2005}, attributing to the material an elastic instantaneous strain together with a retarded dampening effect. This aspect is particularly visible in case of hydraulic transients, for which it has already been demonstrated that the classical Allievi-Joukowsky theory for water hammer, based on the assumption of a linear elastic wall behavior and quasi-steady friction losses \cite{chaudhry}, fails in the prediction of the real pressure trend in flexible tubes \cite{covas2004,covas2005}. From the experimental point of view, a recent thorough work has been done by Ferr\`{a}s et al. \cite{ferras2016} for the distinction of the main effects of dampening during hydraulic transients in PE pipes. Ramos et al. \cite{ramos2004} discussed the importance of the implementation of a visco-elastic constitutive law for plastic pipes and also the relevance of the unsteady friction with respect to the steady one. Their results show that the pressure wave dissipation is more sensitive to the visco-elastic damping effects than to the unsteady friction losses. Furthermore, Duan et al. \cite{duan2010} demonstrated that the visco-elastic effects are deeply more significant when the retardation time is less than the wave travel time along the entire pipeline length. Other researches regarding the unsteady friction losses were already been done by Zielke \cite{zielke1968} and Franke \cite{franke1983}, while recently Ioriatti et al. \cite{IDIzamm} proposed a new more efficient approach for evaluating the convolution integral of the unsteady wall shear stress.\par
In many industrial applications involving the design of hydraulic networks accurate computational models able to correctly a priori evaluate the behaviour of the systems are required. The mathematical model has to properly describe the hydraulic system also in terms of resistance and deformation of the pipe wall, especially in the event of water hammers which could seriously damage the whole system. Moreover, considering the increase in complexity of these systems, numerical simulations have to be more and more efficient and robust \cite{leibinger2016}. The main numerical method used for studies concerning hydraulic transients has always been the Method of Characteristics (MOC) \cite{ghidaoui2005}. Among these studies, a lot of research has been done for the single-pipe plastic system by Covas et al. \cite{covas2004,covas2005}, Soares et al. \cite{soares2008} and Apollonio et al. \cite{apollonio2014}. There are applications carried out also with a 2D axially symmetric model in \cite{duan2010, pezzinga2014}. Meniconi et al. \cite{meniconi2012, meniconi2014} analyzed the effect of water hammer pressure waves in case of sudden contraction or expansion of the cross-sectional area or with an in-line valve in the pipeline. Evangelista et al. \cite{evangelista2015} also investigated the behavior of more complex hydraulic systems, with a Y-shaped configuration. \par Other techniques are only seldom applied for the resolution of transient pipe flows and especially include Finite Volume Methods (FVM) \cite{seck2017}. Starting from this consideration, in the present work we test the Path-Conservative Osher-type Explicit Finite Volume Method (so-called DOT Riemann solver \cite{dumbser2011,dumbser2011a}) and the Semi-Implicit Staggered Finite Volume Method (further simply called SI) presented in \cite{dumbser2015} with two water hammer problems in single HDPE pipelines. Then we compare the results, in terms of accuracy and efficiency, to those obtained applying the classical MOC. It has to be mentioned that the DOT solver had never been used before for this type of applications, only for frequency analysis in \cite{leibinger2016}, while the SI method had already been tested with hydraulic transients, but only considering an elastic tube-wall behaviour \cite{IDIzamm}. In the present research, water hammer test cases are carried out taking into account different linear visco-elastic rheological models: the Standard Linear Solid Model (SLSM) and the generalized Kelvin-Voigt chain, with the aim to evaluate if a more complex model is worth to be chosen for achieving a better agreement with experimental data. To the authors' knowledge, this work is the first one extending the applicability of the generalized Kelvin-Voigt model both to the Explicit and the Semi-Implicit numerical schemes. Furthermore, we made a comparison of the results obtained implementing a quasi-steady friction model and an unsteady friction model, with the approach proposed in \cite{IDIzamm}, applied in case of turbulent flow for the first time in literature. To stress more these numerical schemes in order to reveal their weaknesses, tests have been executed also with three demanding Riemann problems (i.e. initial value problems governed by conservation laws with piecewise constant initial data having a single discontinuity \cite{Toro2009}), adopting an elastic rheological behaviour of the tube wall. The aim of the RP here presented is to evaluate the robustness of each scheme, pointing out the performance of every method in case of cross-sectional changes, when more flexible materials are considered and when cavitation occurs. \par
The paper is structured as follows: in section \ref{S:2} the mathematical model is presented, with the specific characterization made for each numerical scheme and for each constitutive tube law and friction model chosen. In section \ref{S:3} the three numerical models, MOC, DOT and SI, are described. In section \ref{S:5} first the relevance of the unsteady friction model is analysed and the calibration procedure for the visco-elastic parameters is briefly illustrated; then the various test cases are presented, discussing the most interesting results. Finally, in section \ref{S:6}, some comments about the novel comparison of the numerical schemes and their modelling in parallel with numerical results are reported.

\section{Mathematical model}
\label{S:2}
The governing balance laws system of a compressible fluid through a flexible tube is obtained averaging the 3D compressible Navier-Stokes equations over the cross-section under the assumption of axially symmetry of the geometry of the conduct and of the flow \cite{ghidaoui2005}. \\ 
The resulting simplified 1D non-linear hyperbolic system of partial differential equations (PDE) is composed by the continuity equation and the momentum equation and reads \cite{ghidaoui2005}:
\begin{subequations}\label{eq:sysinitial}
\begin{flalign}
\label{eq:continuity}
&\frac{\partial}{\partial t} (A\rho) + \frac{\partial}{\partial x} (A\rho u) = 0\\
\label{eq:momentum}
&\frac{\partial}{\partial t} (A\rho u) + \frac{\partial}{\partial x} (A\rho u^2 + Ap) - p \frac{\partial A}{\partial x} = F_R ,
\end{flalign}
\end{subequations}
where \(x\) is the space, \(t\) is the time, \(A\) is the cross-sectional area, \(\rho\) is the cross-sectional averaged density of the fluid, \(u\) is the averaged fluid velocity, \(p\) is the averaged fluid pressure and \(F_R\) is the source term accounting for the friction between fluid and tube wall, discussed in section \ref{S.S:FR}.\par
To close system \eqref{eq:sysinitial}, an equation of state (EOS) and a tube constitutive law must be added. \\ 
In most of the technical applications it is usually sufficient to assume a barotropic behaviour of the fluid, therefore $\rho = \rho(p)$. Nevertheless, taking into account the cavitation phenomena may be useful. An EOS for barotropic flow which takes into account cavitation is presented in section \ref{S.S:EOS}.\\ 
The tube law, that describes the relationship between the tube cross-section and the internal pressure containing all the information about the rheological behavior of the pipe material, can be expressed in different ways. Since we want to take into account the deformability and the flexibility of the tube wall, in this paper we consider two different rheological models: the first one for characterizing an elastic behavior and the second one for a more complex visco-elastic behavior, which is necessary to reproduce the real performance of plastic tubes \cite{evangelista2015,covas2004,covas2005}. These models are presented in sections \ref{S.S:2.1} and \ref{S.S:2.2}.\par
Returning to system \eqref{eq:sysinitial}, it is possible to derive the classical water hammer equations in terms of piezometric head \(h\) and velocity \(u\), when temperature changes can be neglected.
From Eq.~\eqref{eq:continuity} we obtain:
\begin{equation}
\label{eq:cont.CC}
	\rho \left(\frac{\partial A}{\partial t} + u \frac{\partial A}{\partial x}\right) + A \left( \frac{\partial 	\rho}{\partial t} + u \frac{\partial \rho}{\partial x}\right) + A\rho \frac{\partial u}{\partial x} = 0.
\end{equation}
In parallel, manipulating Eq.~\eqref{eq:momentum}, with $g$ the gravity acceleration and $j$ the frictional head losses per unit length (see section \ref{S.S:FR}), we get:
\begin{equation}
\label{eq:mom.CC}
	\frac{1}{g} \left( \frac{\partial u}{\partial t} + u\frac{\partial u}{\partial x} \right)+ \frac{1}{\rho g} \frac{\partial p}{\partial x} = -j .
\end{equation}
If we make the assumption that $\frac{1}{\rho g} \frac{\partial p}{\partial x} \approx \frac{\partial}{\partial x } \left(\frac{p}{\rho g}\right)$, and therefore that the intrinsic spatial variation of the density is negligible (valid assumption if we are considering a weakly compressible fluid), Eq.~\eqref{eq:mom.CC} becomes
\begin{equation}
\label{eq:mom.CC1}
	\frac{1}{g} \left( \frac{\partial u}{\partial t} + u\frac{\partial u}{\partial x} \right)+ \frac{\partial h}{\partial x} = -j,
\end{equation}
with $h={p}/{\rho g}$.
Let us now introduce a generic function \(\mathcal{F}(x,t)\) that represents any properties of the pressure wave concerning water hammer problems. Standing on the wave frame reference, the property \(\mathcal{F}\) remains constant in both time and space, thus it can be written that:
\begin{equation*}
	\frac{\d \mathcal{F}}{\d t} = \frac{\partial \mathcal{F}}{\partial t} + \frac{\partial \mathcal{F}}{\partial x} \frac{\d x}{\d t} = 0 ,
\end{equation*}
being ${\d x}/{\d t}$ coincident with the pressure wave celerity $c$.
Rearranging this equation and dividing for the velocity \(u\) in the duct, we obtain:
\begin{equation}
\label{eq:c/u}
	\frac{c}{u} = -\frac{\partial \mathcal{F} / \partial t}{u\, \partial \mathcal{F}/ \partial x} .
\end{equation}
Since, in a general water hammer problem in flexible tubes (and even more in rigid tube cases), the speed \(c\) is considerably bigger than $u$ \cite{Wylie1978}, from (\ref{eq:c/u}) follows that all terms \(u\, \partial \mathcal{F}/ \partial x\) can be neglected compared to terms \(\partial \mathcal{F} / \partial t\). Thus, system of Eqs.~\eqref{eq:cont.CC} and \eqref{eq:mom.CC1} becomes:
\begin{subequations}\label{eq:WH}
\begin{flalign}
\label{eq:cont.WH}
	&\rho \frac{\partial A}{\partial t} + A  \frac{\partial \rho}{\partial t} + A\rho \frac{\partial u}{\partial x} = 0\\
\label{eq:mom.WH}
	&\frac{1}{g} \frac{\partial u}{\partial t} + \frac{\partial h}{\partial x} = -j .
\end{flalign}
\end{subequations}

Equation \eqref{eq:cont.WH} can be manipulated in order to be written in terms of piezometric head and velocity. Considering again the assumption $ \frac{1}{\rho g} \frac{\partial p}{\partial t} \approx \frac{\partial h}{\partial t}$, we can write: 
\begin{subequations}\label{eq:dh}
\begin{flalign}
\label{eq:dh.A}
	&\frac{\partial A}{\partial t} = \frac{\partial A}{\partial p}\frac{\partial p}{\partial t} = \rho g \frac{\partial A}{\partial p}\frac{\partial h}{\partial t}\\
\label{eq:dh.rho}
	&\frac{\partial \rho}{\partial t} = \frac{\partial \rho}{\partial p}\frac{\partial p}{\partial t} = \rho g \frac{\partial \rho}{\partial p}\frac{\partial h}{\partial t}.
\end{flalign}
\end{subequations}
Substituting Eq.~\eqref{eq:dh} into \eqref{eq:cont.WH}, we get the simplified unsteady pipe flow system of equations, in which the convective transport terms have been neglected (classical Allievi-Joukowsky theory):
\begin{subequations}\label{eq:WH1}
\begin{flalign}
\label{eq:cont.WH1}
	&\frac{\partial h}{\partial t} + \frac{ c^2}{g} \frac{\partial u}{\partial x} = 0  \\
\label{eq:mom.WH1}
	&\frac{1}{g} \frac{\partial u}{\partial t} + \frac{\partial h}{\partial x} = -j,
\end{flalign}
\end{subequations}
where the celerity $c$ is given by:
\begin{equation}
\label{eq:celerity}
c = \sqrt{\frac{\frac{\partial p}{\partial \rho}}{1+\frac{\rho}{A}\frac{\partial A}{\partial p}\frac{\partial p}{\partial \rho}}} = \frac{c'_0}{\sqrt{1+ \frac{\rho c_0'^2}{A \beta'}}},
\end{equation}
with $c_0'(p) = \sqrt{\partial p/\partial \rho}$ and $\beta'(p)= \partial p/\partial A$. \\ 
Equation \eqref{eq:celerity} can be made explicit when suitable tube law and EOS are selected.

\subsection{Equation of state}
\label{S.S:EOS}
Assuming a barotropic fluid, the density only depends on the pressure, hence \(\rho = \rho(p)\). Taking also into account cases in which cavitation occurs, and thus supposing to have cases in which $p < p_v$, where $p_v$ is the vapor pressure, the following equation of state with an homogeneous mixture approximation is selected \cite{dumbser2015}:
\begin{equation}
\label{eq:eos}
    \rho(p)=\left\{
                \begin{array}{ll}
                \rho_0 + \frac{1}{c_0^{2}} (p-p_v) \quad &\text{if} \quad p \geq p_v\\
                \left[\frac{\varphi(p)}{\rho_v(p)}+\frac{1-\varphi(p)}{\rho_0}\right]^{-1} \quad &\text{if} \quad 0<p<p_v\\
                \end{array}
              \right.
\end{equation}
with \(\rho_0\) and \(p_0\) the reference density and pressure in equilibrium state respectively, $c_0$ the speed of sound in the fluid at reference conditions, \(\varphi(p) = -K(p-p_v)\) the mass fraction of vapour, with \(K\) cavitation constant, and \(\rho_v(p) = \frac{p}{R_v T_0}\) the vapour density, calculated considering the gas constant \(R_v\) and the reference temperature \(T_0\). 

\subsection{Friction model}
\label{S.S:FR}
Concerning the friction model applied to water hammer problems and introducing the ratio of the diffusion time scale to the wave time scale:
\begin{equation}
\label{eq:P}
	\P = \frac{2D/f u_0}{L/c}  ,
\end{equation}
where $D$ is the pipe diameter, $L$ is the length of the pipe, $f$ is the friction factor, as defined by the Darcy-Weisbach formula \cite{Wylie1978}, $u_0$ is the initial velocity and $c$ is given by Eq.~\eqref{eq:celerity}, it has been shown that accurate physically based unsteady friction models are required if $\P$ is of order 1 or less \cite{ghidaoui2002,duan2010}.


If $\P \gg 1$, it is possible to consider only a quasi-steady friction model, for which the term \(F_R\) in (\ref{eq:momentum}) reads:
\begin{equation}
\label{eq:FR}
	F_R = -A\rho g j,
\end{equation}
with the frictional head loss per unit length $ j = \frac{f}{D} \frac{u| u|}{2g}$. Considering a cylindrical tube with axially symmetric flow, the same quantity \(F_R\) can also be expressed in terms of the wall shear stress \(\tau_w\) as:
\begin{equation}
\label{eq:FR2}
	F_R = -2\pi R\tau_w,
\end{equation}
where $R=D/2$ is the pipe radius and with \(\tau_w\) concerning only the quasi-steady contribute, \(\tau_s\), hence:
\[\tau_w = \tau_s = f\frac{\rho u |u|}{8} .\]

If it is necessary to take into account also unsteadiness effects, i.e. $\P \ll 1$, the wall shear stress \(\tau_w\) must be written as sum of quasi-steady, $\tau_s$, and unsteady stresses, $\tau_u$:
\begin{equation}
\label{eq:tau}
	\tau_w = \tau_s + \tau_u
\end{equation}
and thus, considering the expression of Zielke \cite{zielke1968}:
\begin{equation}
\label{eq:tau_zielke}
	\tau_w = f\frac{\rho u |u|}{8} + \frac{2\mu}{R}\int_0^t  w(t-t')\frac{\partial u}{\partial t}(t')\,\d t' ,
\end{equation}
where \(\mu\) is the dynamic viscosity, \(w\) is a weighting function and \(t'\) is the integral variable having dimension of time.\par
The evaluation of the convolution integral in Eq.~\eqref{eq:tau_zielke} is very time consuming and several solutions have been proposed (see \cite{Urbanowicz,shamloo2015} for extensive summaries). The first researcher who developed an effective method is Trikha \cite{trikha1975}, while the most diffused formulation is the one proposed by Kagawa \cite{Kagawa}, who improved Trikha's approach. Recently a novel approach has been proposed in \cite{IDIzamm} in the case of laminar flow. In particular, the solution of the convolution integral is reduced to the solution of a set of ordinary differential equations (ODEs). This allows to gain efficiency with respect the formula of Kagawa (which requests the evaluation of exponential functions) and in the following we extend this last approach, proposing the ODE Model for turbulent flow cases. First, we consider the weighting function in turbulent regime proposed by Urbanowicz and Zarzycki \cite{Urbanowicz}, expressed as a series of exponential functions:
\begin{equation}
w(t)=\sum_{i=1}^{N_w}A^*m_i^*\exp\left[{-(n_i^*+B^*)\frac{\nu t}{R^2}}\right],
\label{wapp}
\end{equation}
where $\nu$ is the cinematic viscosity and $N_w=16$, $ (n_1^*,...,n_{16}^*)=$ (4.78793, 51.0897, 210.868, 765.03, 2731.01, 9731.44, 34668.5, 123511, 440374, 1578229, 5481659, 18255921, 59753474, 192067361, 616415963, 1945566788) and
$(m_1^*,...,m_{16}^*)=$ (5.03392, 6.4876, 10.7735, 19.904, 37.4754, 70.7117, 133.460, 251.933, 476.597, 902.22, 1602.04, 2894.84, 5085.55, 9190.11, 16118.6, 29117.3). Moreover, we will consider only smooth pipes in our simulation, so the parameters $A^*$ and $B^*$ are those proposed by Vardy and Brown \cite{vardy2003}:
\begin{equation}
A^*=\sqrt{1/(4\pi)} \quad B^*=\Re^k/12.86 \quad \text{with} \quad k=	\log_{10}(15.29/\Re^{0.0567}),
\end{equation}
where $\Re={|u| D}/{\nu}$ is the Reynolds number.
Substituting  Eq.~\eqref{wapp} into the convolution integral, the unsteady wall shear stress is computed as:
\begin{equation}
\tau_u=\sum_{i=1}^{N_w} \tau_i= \sum_i^{N_w} \frac{2\mu}{R} \int_0^t A^* m_i^* \exp\left[-\frac{\nu(n_i^*+B^*)}{R^2}(t-t')\right] \frac{\d u}{\d t}(t')\, \d t'.
\end{equation}
Then, the left and the right side of the \textit{i}-th contribution in the last equation are derived with respect to the time. Applying the Leibniz rule yields the following ODE:
\begin{equation}
\frac{\d}{\d t}\tau_i = - \frac{(n_i+B^*) \nu}{R^2} \tau_i + \frac{2 \mu}{R}\frac{\d u }{\d t}m_i^*A^*,
\end{equation}
which is discretised in time using the implicit Euler method (see \cite{IDIzamm}):
\begin{equation}
\frac{\tau_i^{n+1}-\tau_i^{n}}{\Delta t} = - \frac{(n_i+B^*) \nu}{R^2} \tau_i^{n+1} + \frac{2 \mu}{R}\frac{u^{n+1}-u^{n}}{\Delta t}m_i^*A^*.
\end{equation}
Finally, the total unsteady wall shear stress at the time $t ^{n+1} = t^n+\Delta t$ is computed as follows: 
\begin{equation}
\tau_u^{n+1}=\sum_{i=1}^{N_w}\tau_i^{n+1} \quad \text{with} \quad \tau_i^{n+1}=\frac{\tau_i^n+\frac{2\mu}{R}(u^{n+1}-u^n)m_i^*A^*}{1+\frac{(n_i+B^*) \nu}{R^2}\Delta t} \quad i=1,2,... N_w \quad.
\end{equation}

\subsection{Elastic constitutive tube law}
\label{S.S:2.1}
For studying water hammer events occurring in commercial pipes is usually sufficient to consider an elastic rheological behaviour of the tube wall, in particular when dealing with steel ducts, but also to obtain a first fair approximation working with plastic pipes.
First we consider Hooke's law \cite{avallone1916}:
\begin{equation}
\label{eq:hooke2}
	\d\sigma = E_0 \frac{\d D}{D},
\end{equation}
with \(E_0\) instantaneous Young (elastic) modulus of the material and Barlow's formula \cite{avallone1916}:
\begin{equation}
\label{eq:barlow}
	pD = 2\sigma s ,
\end{equation}
with \(s\) thickness of the tube and \(\sigma\) traction tension; with suitable manipulations, integrating and linearizing the combination of the two equations, it is possible to obtain the so called Laplace law \cite{Wylie1978}:
 \begin{equation}
\label{eq:laplace}
	A = A_0 + \frac{1}{\beta} (p-p_0),
\end{equation}
where \(A_0\) is the equilibrium cross-sectional area of the tube, related to the equilibrium pressure \(p_0\).
In this elastic constitutive equation the cross-sectional area \(A\) only depends linearly on the pressure \(p\) through a coefficient \(\beta\), that contains all the elastic properties of the material:
\begin{equation}
\label{eq:beta}
	\beta = \frac{E_0 s}{D_0 A_0} = \frac{\sqrt{\pi} E_0 s}{2 A_0 \sqrt{A_0}},
\end{equation}
with \(D_0\) the diameter that corresponds to the equilibrium area \(A_0\).\par

As suggested in \cite{leibinger2016}, to get the closure equation for system \eqref{eq:sysinitial}
, we need to differentiate Eq.~\eqref{eq:laplace} with respect to \(t\) and to recur to Eq.~\eqref{eq:continuity}, thus:
\begin{equation}
\label{eq:laplace.diff}
	\frac{\partial A}{\partial t} + \frac{1}{\frac{\beta A}{c_0'^2} + \rho} \frac{\partial (A\rho u)}{\partial x} = 0 ,
\end{equation}
where $c_0'(p) = \sqrt{\partial p/\partial \rho}$ depends on the selected EOS.

\subsection{Visco-elastic constitutive tube laws}
\label{S.S:2.2}
\begin{figure}[t]
\centering
	\begin{subfigure}[b]{0.45\textwidth}
	\centering\includegraphics[width=0.75\linewidth]{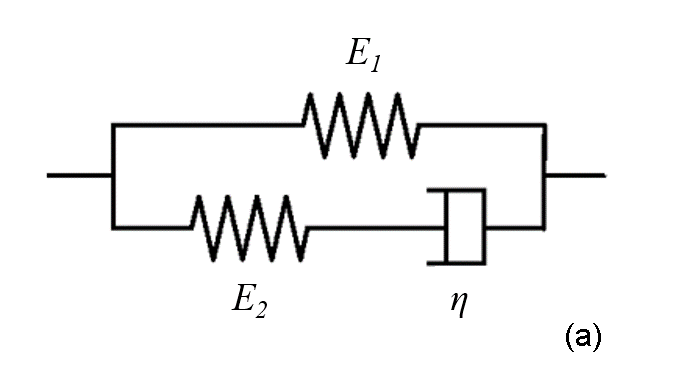}
	\label{fig.3parMaxwell}
	\end{subfigure}
	\begin{subfigure}[b]{0.45\textwidth}
	\centering\includegraphics[width=0.75\linewidth]{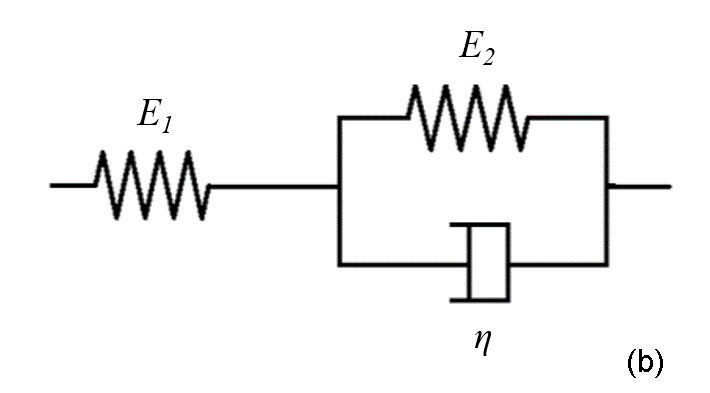}
	\label{fig.3parKelvin}
	\end{subfigure}
\caption{Schemes of the 3-parameter visco-elastic SLSM, (a) with Maxwell unit and (b) with Kelvin-Voigt unit.}
\label{fig.3par}
\end{figure}

To better reproduce the real behaviour of a polymer material, as those used in industries, it is necessary to introduce a visco-elastic model. \\
A constitutive relation of linear visco-elasticity is built up, considering the material as a sum of linear elements such as (elastic) linear springs and (viscous) linear dash pots, to take into account also the time dependent relaxation of the wall.\\
The simplest model able to correctly reproduce the stress-strain behavior of a polymer material is the 3-parameter model known as Standard Linear Solid Model (SLSM), which can be structured with a Maxwell or a Kelvin-Voigt element \cite{Lakes2009}. The schematic representations of both the types are shown in Fig.~\ref{fig.3par}. \par
To obtain more flexible models, it is possible to extend the chain of Maxwell or Kelvin-Voigt elements to an infinite number. Theoretically, the more elements we have, the more accurate our model will be in describing the real response of the material. Conversely, the more complex the model is, the more material parameters that must be calibrated there are.
For further information about visco-elasticity models, the reader can refer to \cite{gurtin1962,Lakes2009}. \par
However, a sensitivity analysis was carried out in \cite{covas2005} to estimate the number of Kelvin-Voigt elements beyond which the accuracy of the results doesn't improve anymore, and the optimal number resulted equal to 4. \\
For the comparisons presented in this paper, we first consider the 3-parameter model and, to test the extension to more complex multi-parameter models, we carry out some tests also considering 5-parameter models, as the one used in \cite{evangelista2015}.\par
Concerning the two SLSM, 3-parameter Maxwell model and 3-parameter Kelvin-Voigt model, it can be demonstrated that they reproduce exactly the same behaviour of the material. As a matter of fact, if we evaluate the constitutive equation of the 3-parameter model, between stress \(\sigma\) and strain \(\epsilon\),
\begin{equation}
\label{eq:3par}
	\frac{\d\sigma}{\d t} = E_0 \frac{\d\epsilon}{\d t} - \frac{1}{\tau_r}(\sigma - E_\infty \epsilon),
\end{equation}
we have that in the Maxwell model case (Fig.~\ref{fig.3par}a), the instantaneous Young modulus \(E_0\), the asymptotic Young modulus \(E_\infty\) and the relaxation time \(\tau_r\) are:
\[E_0 = E_1 + E_2 , \quad E_\infty = E_1, \quad \mathrm{and} \quad \tau_r = \frac{\eta}{E_2} ;\]
while for the Kelvin-Voigt model case (Fig.~\ref{fig.3par}b):
\[E_0 = E_1 , \quad E_\infty = \frac{E_1E_2}{E_1 + E_2} \quad \mathrm{and} \quad \tau_r = \frac{\eta}{E_1 + E_2} .\] \par
To obtain the visco-elastic material closing equation for system \eqref{eq:sysinitial} we consider the procedure presented in \cite{leibinger2016} and applying Barlow's formula we have:
\begin{equation}
\label{eq:3par.diff}
	\frac{\partial A}{\partial t} + \frac{1}{\frac{E_0\phi}{2W c_0'^2} + \rho} \frac{\partial (A\rho u)}{\partial x} = \frac{[2W(p-p_0) - E_\infty(\phi-1)]A}{\tau_r(2Wc_0'^2\rho + E_0 \phi)},
\end{equation}
that is the constitutive partial differential equation of the SLSM, where the parameter \(W = \alpha D/2s\) is the ratio between the radius and the wall thickness of the pipe multiplied by the axial pipe-constraint dimensionless parameter \(\alpha = \frac{2s}{D}(1+\nu_p)+\frac{D}{D+s}(1-\nu_p^2)\) for a wall pipe anchored along its length (see \cite{Wylie1978}); \(\nu_p\) is the Poisson's ratio and \(\phi = A/A_0 = (1+\epsilon)^2 \approx 1+2\epsilon\) is the normalized cross-sectional area. Comparing this equation to the Laplace PDE (\ref{eq:laplace.diff}), it can be noticed that all the viscous properties of the material are contained in the source term.\par
Furthermore, it is possible to write the PDE (\ref{eq:3par.diff}) as an ODE.
In fact, multiplying by the term $[1+ 2 W c_0'^2 \rho/(\phi E_0)]$ and recurring to some algebraic manipulations, Eq.~\eqref{eq:3par.diff} becomes:
\begin{equation}
\label{eq:3par.diff_1}
	\frac{\partial A}{\partial t} = -\frac{2Wc_0'^2}{E_0} \frac{A_0}{A} \left[ \frac{\partial (A\rho u)}{\partial x} +\rho \frac{\partial A}{\partial t}\right] + \frac{2WA_0(p-p_0)}{\tau_r E_0} - \frac{(A-A_0)E_\infty}{\tau_r E_0} .
\end{equation}
Indicating again with $c_0'(p) = \sqrt{\partial p/\partial \rho}$ the speed of sound related to the selected  EOS, we observe that:
\begin{equation}
\label{eq:dp/dt}
	\frac{\partial p}{\partial t} = \frac{\partial p}{\partial \rho} \frac{\partial \rho}{\partial t} = c_0'^2 \frac{\partial \rho}{\partial t}.
\end{equation}
At the same time,
\[\frac{\partial (A\rho)}{\partial t} = A \frac{\partial \rho}{\partial t} + \rho\frac{\partial A}{\partial t},\]
thus,
\begin{equation}
\label{eq:drho/dt}
	\frac{\partial \rho}{\partial t} = \frac{1}{A} \left[\frac{\partial (A\rho)}{\partial t} - \rho\frac{\partial A}{\partial t}\right] .
\end{equation}
From (\ref{eq:dp/dt}) and (\ref{eq:drho/dt}), using the continuity Eq.~\eqref{eq:continuity}, we get:
\begin{equation}
\label{eq:dp/dt_1}
	\frac{\partial p}{\partial t} = - \frac{c_0'^2}{A} \left[\frac{\partial (A\rho u)}{\partial x} + \rho\frac{\partial A}{\partial t}\right] .
\end{equation}
If we use Eq.~\eqref{eq:dp/dt_1} into Eq.~\eqref{eq:3par.diff_1}, the ODE results:
\begin{equation}
\label{eq:3par.ODE}
	\frac{\d A}{\d t} = \frac{2WA_0}{E_0} \frac{\d p}{\d t} + \frac{2WA_0(p-p_0)}{\tau_r E_0} - \frac{(A-A_0)E_\infty}{\tau_r E_0} .
\end{equation}\par

\begin{figure}[t]
\centering\includegraphics[width=0.55\linewidth]{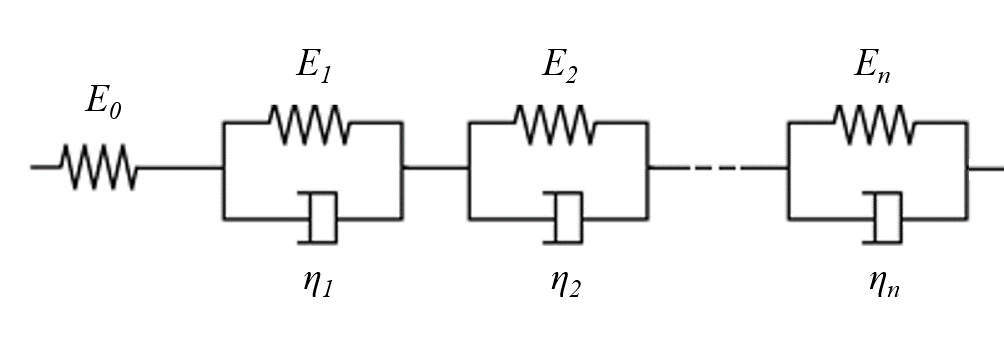}
\caption{Scheme of the generalized Kelvin-Voigt chain as multi-parameter visco-elastic model.}
\label{fig.kelvinchain}
\end{figure}

To extend the applicability of Eq.~\eqref{eq:3par.diff} to visco-elastic models that include more than three parameters, an original formulation is here presented for the first time. We consider a generalized Kelvin-Voigt chain, with \(N_{KV}\) Kelvin-Voigt (KV) units in series and one isolated spring, as shown in Fig.~\ref{fig.kelvinchain}. 
With this model system we obtain the following equations for the basic stress-strain relations (with subscript \(k\) referring to parameters of the \(k^{th}\) element):
\[ \epsilon(t) = \epsilon_0 + \sum_{k=1}^{N_{KV}} \epsilon_{rk}\]
\[ \sigma(t) = \sigma_0 (t) = \sigma_k (t) \]
\[\sigma_k (t) = \sigma_{k,E} + \sigma_{k,D} = E_k \epsilon_{rk} + \eta_k \frac{\partial \epsilon_{rk}}{\partial t} .\]
From which follows:
\[\frac{\partial \epsilon}{\partial t} = \frac{\partial \epsilon_0}{\partial t} + \sum_{k=1}^{N_{KV}} \frac{\partial \epsilon_{rk}}{\partial t} = \frac{1}{E_0} \frac{\partial \sigma}{\partial t} + \sum_{k=1}^{N_{KV}} \left( \frac{\sigma - E_k\epsilon_{rk}}{\eta_k} \right) \]
and thus,
\begin{equation}
\label{eq:npar}
	\frac{\partial \epsilon}{\partial t} = \frac{1}{E_0} \frac{\partial \sigma}{\partial t} + \sigma \sum_{k=1}^{N_{KV}} \frac{1}{\eta_k}  - \sum_{k=1}^{N_{KV}} \frac{E_k\epsilon_{rk}}{\eta_k} .
\end{equation} 
Recurring again to Barlow's formula \eqref{eq:barlow}, with the same procedure mentioned above for the 3-parameter case, we finally obtain the closure equation for the generalized Kelvin-Voigt model:
\begin{equation}
\label{eq:npar.diff}
	\frac{\partial A}{\partial t} + \frac{1}{\frac{E_0\phi}{2W c_0'^2} + \rho} \frac{\partial (A\rho u)}{\partial x} = \frac{2E_0 A\left[W(p-p_0) \sum_{k=1}^{N_{KV}} \frac{1}{\eta_k} - \sum_{k=1}^{N_{KV}} \frac{\epsilon_{rk}}{\tau_{rk}} \right]}{2Wc_0'^2\rho + E_0 \phi}.
\end{equation}\par
With the same procedure presented above for the 3-parameter model, it is possible to obtain the related ODE of \eqref{eq:npar.diff}, thus:
\begin{equation}
\label{eq:npar.ODE}
	\frac{\d A}{\d t} = \frac{2WA_0}{E_0} \frac{\d p}{\d t} + 2A_0\left[W(p-p_0) \sum_{k=1}^{N_{KV}}\frac{1}{\eta_k} - \sum_{k=1}^{N_{KV}}\frac{\epsilon_{rk}}{\tau_{rk}} \right].
\end{equation}\par

To consider the system of the classical water-hammer equations taking into account the visco-elasticity of the tube wall, we need to add a specific term to the continuity Eq.~\eqref{eq:cont.WH1}, whereas the momentum Eq.~\eqref{eq:mom.WH1} remains unaltered:
\begin{equation}
\label{eq:cont.WHvisco}
	\frac{\partial h}{\partial t} + \frac{ c^2}{g} \frac{\partial u}{\partial x} = - \frac{2c^2}{g} \frac{\d\epsilon_r}{\d t},
\end{equation}
 with \(\epsilon_r\)  representing the retarded strain \cite{covas2003}.\par
Boltzmann superposition principle \cite{shaw2005} states that, for small strains, each increment of load makes an independent and linearly additive contribution to the total deformation. Thus, the elastic deformation is given by: 
\begin{equation}
\label{eq:epse}
\epsilon_e(t) = J_0\sigma(t),
\end{equation}
in which \(J_0\) is the instantaneous creep compliance (equal to the inverse of the instantaneous modulus of elasticity \(E_0\) for linear visco-elastic materials), while the retarded deformation can be written as:
\begin{equation}
\label{eq:epsir}
\epsilon_r(t) = \int_{0}^{t} \sigma(t - t') \frac{\partial J}{\partial t}(t')\,\d t',
\end{equation}
with $J(t')$ the creep function at time $t'$. Finally the total deformation will be the sum of the two, \(\epsilon(t) = \epsilon_e(t) + \epsilon_r (t)\). By applying Barlow's formula, considering also the axial pipe-constraint dimensionless parameter \(\alpha\), the total circumferential strain can also be expressed as (see \cite{covas2005}):
\begin{equation}
\label{eq:totalstrain.CC}
	\epsilon (t) = \epsilon_e + \epsilon_r = \frac{\alpha D_0}{2s_0}[p(t) - p_0]J_0 + \int_{0}^{t} \frac{\alpha D(t-t')}{2s(t-t')}[p(t-t')-p_0]\frac{\partial J}{\partial t}(t') \d t'.
\end{equation}
This equation is valid for the selected number of Kelvin-Voigt elements, being the creep function of the pipe wall represented by a mathematical expression that can be implemented numerically,
\begin{equation}
\label{eq:sumJ}
	J(t) = J_0 + \sum_{k=1}^{N_{KV}} J_k \left(1-e^{-\frac {t}{\tau_{rk}}}\right) ,
\end{equation}
with \(N_{KV}\) Kelvin-Voigt elements, \(J_k = 1/E_k\) and \(\tau_{rk} = \eta_k/E_k\) the visco-elastic parameters of the \(k^{th}\) Kelvin-Voigt element. It is worth noting that for the SLSM of Kelvin-Voigt type $N_{KV}$ is equal to 1. \par

\subsection{Complete coupled systems of the FSI problem}
\label{S.S:2.3}
As described in the previous sections, the continuity and the momentum equations can be expressed in different forms. Moreover, the closure equations have different formulations depending on the assumed behaviour of the pipe material and the relevance of the flow unsteadiness in the computation of the friction effects. 
The more appropriate formulation of the system of governing equations depends on the chosen numerical integration technique (MOC, DOT or SI). In this section, we summarize, for each numerical method considered in this work, the most suited form of the complete system of equations.

To take into account the fluid-structure interaction (FSI), working in the context of the Explicit Finite Volume Method \cite{leibinger2016}, the PDE of the material model have to be added to the system of averaged Navier-Stokes equations \eqref{eq:sysinitial}, obtaining the system:
\begin{subequations}\label{eq:sysDOT}
\begin{flalign}
\label{sysDOT.1}
	&\frac{\partial}{\partial t} (A\rho) + \frac{\partial}{\partial x} (A\rho u) = 0 \\
\label{sysDOT.2}
	&\frac{\partial}{\partial t} (A\rho u) + \frac{\partial}{\partial x} (A\rho u^2 + Ap)  - p \frac{\partial A}{\partial x} = - 2\pi R \tau_w\\
\label{sysDOT.3}
	&\frac{\partial}{\partial t} A +d \frac{\partial}{\partial x} (A\rho u)= S \\
\label{sysDOT.4}
	&\frac{\partial}{\partial t} A_0 = 0 .
\end{flalign}
\end{subequations}
Equation \eqref{sysDOT.3} unifies both the elastic and the visco-elastic wall models \cite{leibinger2016}. Concerning the Laplace elastic law:
\begin{equation}
\label{dS_laplace}
	d = \frac{1}{\frac{\beta A}{c_0'^2}+ \rho} \qquad \mathrm{and} \qquad S = 0,
\end{equation}
for the visco-elastic 3-parameter model:
\begin{equation}
\label{dS_3par}
	d =\frac{1}{\frac{E_0\phi}{2W c_0'^2} + \rho} \qquad \mathrm{and} \qquad S = \frac{[2W(p-p_0) - E_\infty(\phi-1)]A}{\tau_r(2Wc_0'^2\rho + E_0 \phi)} 
\end{equation}
and in general for the visco-elastic multi-parameter model:
\begin{equation}
\label{dS_multipar}
	d =\frac{1}{\frac{E_0\phi}{2W c_0'^2} + \rho} \qquad \mathrm{and} \qquad S = \frac{2E_0 A\left[W(p-p_0) \sum_{k=1}^{N_{KV}} \frac{1}{\eta_k} - \sum_{k=1}^{N_{KV}} \frac{\epsilon_{rk}}{\tau_{rk}} \right]}{2Wc_0'^2\rho + E_0 \phi}. 
\end{equation}
The last equation \eqref{sysDOT.4} simply states that the spatially variable equilibrium cross-section $A_0$ is constant in time. This trivial equation is introduced to allow a formally correct treatment of discontinuous longitudinal changes of the reference cross-section $A_0$. In fact, in case of discontinuous $A_0$, the system of governing equations is non-conservative and appropriate numerical techniques must be selected, as done in other contexts \cite{muller2013,gallardo2007}. The Explicit scheme \cite{leibinger2016} belongs to the family of the path-conservative schemes that are specifically developed to address the problem of discontinuous variables arising in applications governed by non-conservative balance laws. The reader is addressed to \cite{Pares2006} for the theory related to the path-conservative schemes.\par

Considering the Semi-Implicit numerical scheme \cite{dumbser2015} for the resolution of the problem, we can more easily consider the two-equation system
\begin{subequations}\label{eq:sysimp}
\begin{flalign}
\label{sysimp.1}
	&\frac{\partial}{\partial t} (A\rho) + \frac{\partial}{\partial x} (A\rho u) = 0 \\
\label{sysimp.2}
	&\frac{\partial}{\partial t} (A\rho u) + \frac{\partial}{\partial x} (A\rho u^2) = - A \frac{\partial p}{\partial x} - 2\pi R \tau_w,
\end{flalign}
\end{subequations}
associated with the tube law expressed by the algebraic equation \eqref{eq:laplace} for an elastic wall model and the ODEs \eqref{eq:3par.ODE} and \eqref{eq:npar.ODE} for the visco-elastic 3-parameter and multi-parameter models, respectively.\par

On the other hand, using the MOC for the discretisation, the classical water hammer equations considering FSI, have the following final form:
\begin{subequations}\label{eq:sysCC}
\begin{flalign}
\label{sysCC.1}
	&\frac{\partial h}{\partial t} + \frac{c^2}{g} \frac{\partial u}{\partial x} = S_M\\
\label{sysCC.2}
	&\frac{1}{g} \frac{\partial u}{\partial t} + \frac{\partial h}{\partial x} = -\frac{4 \tau_w}{\rho g D} = -j .
\end{flalign}
\end{subequations}
with $S_M = 0$ for the elastic wall behaviour and
\begin{equation}
\label{S_VE}
	S_M = - \frac{2c^2}{g} \frac{\d\epsilon_r}{\d t}
\end{equation}
for the visco-elastic wall behaviour. The retarded deformation $\epsilon_r$ in Eq.~\eqref{S_VE} is computed by \eqref{eq:epsir}, selecting the appropriate creep function \eqref{eq:sumJ} for the the 3-parameter and multi-parameter models.\\
Finally, in Eqs.~\eqref{sysDOT.2}, \eqref{sysimp.2} and \eqref{sysCC.2} the wall shear stress $\tau_w$ is computed as $\tau_w=f \rho u|u|/8$ if the quasi-steady model is applicable or using Eq.~\eqref{eq:tau_zielke} if taking into account the effects of unsteadiness of the flow is necessary. 

\section{Numerical models}
\label{S:3}
For solving the mathematical models presented in the previous section \ref{S:2}, three different numerical schemes have been chosen and compared. \\
The standard procedure to solve the simplified system (\ref{eq:sysCC}) in case of water hammer problems is the Method of Characteristics (MOC). 
Other two methods have been tested and compared, in terms of accuracy and efficiency, to the classical MOC: the Explicit Path-Conservative FVM associated with the DOT Riemann solver proposed by Dumbser and Toro in \cite{dumbser2011, dumbser2011a} and the Semi-Implicit (SI) FVM for axially symmetric compressible flows in compliant tubes presented in \cite{dumbser2015}. 

\subsection{Method of Characteristics (MOC)} 
\label{S:3.1}
The simplified system \eqref{eq:sysCC}, obtained by neglecting the convective terms and thus leading to approximately straight characteristic lines \(\Delta x/\Delta t = \pm c\), can be solved by the numerical scheme already presented in \cite{covas2005}:
\begin{equation}
\label{numsys_CC}
	h_i^{n+1}  - h_{i \mp 1}^n \pm \frac{c}{g} \left( u_i^{n+1} - u_{i \mp 1}^n \right) + \frac{2c^2\Delta t}{g} \left(\frac{\d \epsilon_r}{\d t}\right) \pm \frac{f \Delta x}{2gD} u_{i \mp 1}^n \left \lvert u_{i \mp 1}^n\right \rvert = 0
\end{equation}
valid along the characteristic lines, using an uniform grid of \(N_x\) elements with mesh spacing \(\Delta x = x_{i+1}-x_i\) and a time step size \(\Delta t = t^{n+1}-t^{n}\) that respects the \(\CFL\) condition \cite{Toro2009}:
\begin{equation}
\label{eq:cfl}
	\Delta t = \CFL \frac{\Delta x}{max |u \pm c|} .
\end{equation}\\
Thus we get:
\begin{subequations}\label{eq:sysCC_hu}
\begin{flalign}
	&u_i^{n+1} = \frac{u_{i-1}^n + u_{i+1}^n}{2} + \frac{c}{g}\left(\frac{h_{i-1}^n - h_{i+1}^n}{2}\right) - \frac{g\Delta x}{2c}\left(j_{i-1}^n + j_{i+1}^n\right), \\
	&h_i^{n+1} = \frac{h_{i-1}^n + h_{i+1}^n}{2} + \frac{g}{c}\left(\frac{u_{i-1}^n - u_{i+1}^n}{2}\right) - \frac{\Delta x}{2}\left(j_{i-1}^n - j_{i+1}^n\right) - \frac{2c\Delta x}{g} \left(\frac{\d \epsilon_r^n}{\d t}\right) ,
\end{flalign}
\end{subequations}
 with 
\[ j_{i-1}^n = \frac{f}{2gD}u_{i-1}^n \left \lvert u_{i-1}^n\right \rvert \qquad \mathrm{and} \qquad 
j_{i+1}^n = \frac{f}{2gD}u_{i+1}^n \left \lvert u_{i+1}^n\right \rvert.\]
Starting from Eq.~\eqref{eq:totalstrain.CC}, the time-derivative of the viscous retarded strain \(\epsilon_r\) is calculated as a sum of each \(k^{th}\) Kelvin-Voigt element contribution at time \(n\):
\begin{equation}
\label{eq:depsilon_r}
	\frac{\d \epsilon_r^n}{\d t} = \sum_{k=1}^{N_{KV}}\frac{\d \epsilon_{rk}^n}{\d t} = 
	\sum_{k=1}^{N_{KV}} \left[ \frac{\alpha D}{2s} \frac{J_k}{\tau_{rk}} \rho g \left(h^n - h_0\right) - \frac{\epsilon_{rk}^n}{\tau_{rk}}\right],
\end{equation}
considering the numerical approximation of each retarded strain in each node as:
\begin{equation}
\label{eq:numepsilon_r}
	\epsilon_{rk}^n \approx \tilde \epsilon_{rk}^n = J_k F^n - J_k e^{-\Delta t/\tau_{rk}} F^{n-1} - J_k\tau_{rk}\left(1-e^{-\Delta t/\tau_{rk}}\right) \frac{F^n - F^{n-1}}{\Delta t} + e^{-\Delta t/\tau_{rk}} \tilde \epsilon_{rk}^{n-1},
\end{equation}
with the function \(F\) at time \(n\) defined by:
\[F^n = \frac{\alpha D}{2s} \frac{J_k}{\tau_{rk}} \rho g \left(h^n - h_0\right).\]
For further details about this scheme the reader can refer to \cite{covas2005}.\par
When characteristic lines cannot be considered straight, but curves that represent the equation  \(dx/dt = u\pm c\) and the solution is searched in specified intervals both in space and in time (such as for the Riemann problem test cases presented in section \ref{S:5.2}), a linear interpolation from the known values in the grid nodes at each time step needs to be applied \cite{Wylie1978}.

\subsection{Explicit Path-Conservative Finite Volume Method (DOT)}
\label{S:3.2}
A non-linear hyperbolic system of PDE with a conservative and a non-conservative part can be written in the following general form:
\begin{equation}
\label{numsys_gen}
	\frac{\partial \boldsymbol{Q}}{\partial t} + \frac{\partial}{\partial x} \boldsymbol{f}	(\boldsymbol{Q}) + \boldsymbol{B}(\boldsymbol{Q})\frac{\partial \boldsymbol{Q}}{\partial x} = \boldsymbol{S}(\boldsymbol{Q})
\end{equation}
where \(\boldsymbol{Q}\) is the vector of the conserved variables, \(\boldsymbol{f}\) is the flux vector related to the conservative part, \(\boldsymbol{B}(\boldsymbol{Q})\) is the matrix related to the non-conservative part and \(\boldsymbol{S}(\boldsymbol{Q})\) is the source term vector that contains all the head losses and material viscosity information, depending on the visco-elastic model adopted (referring to section \ref{S.S:2.3}). In case the multi-parameter model is chosen, the stress \(\epsilon_{rk}\) of Eq.~\eqref{dS_multipar} can be calculated with the numerical approximation \eqref{eq:numepsilon_r} already presented for the MOC. \\
System \eqref{numsys_gen} can also be written in the quasi-linear form:
\begin{equation}
\label{numsys_quasi-linear}
	\frac{\partial \boldsymbol{Q}}{\partial t} + \boldsymbol{A}(\boldsymbol{Q})\frac{\partial \boldsymbol{Q}}{\partial x} = \boldsymbol{S}(\boldsymbol{Q})
\end{equation}
in which the matrix \(\boldsymbol{A}(\boldsymbol{Q}) = \partial \boldsymbol{f}/ \partial \boldsymbol{Q} + \boldsymbol{B}(\boldsymbol{Q})\) is diagonalizable, with a diagonal matrix \(\boldsymbol{\Lambda}(\boldsymbol{Q})\) containing all real eigenvalues \(\lambda_i\) and a complete set of linearly independent eigenvectors \(\boldsymbol{R}(\boldsymbol{Q})\) (see \cite{leibinger2016}).\par
Indicating with $Q_j$ the $j^{th}$ component of the vector $\boldsymbol{Q}$, for the specific system (\ref{eq:sysDOT}) we have:
\[\boldsymbol{Q} =
\begin{pmatrix} 
  	A\rho \\ A\rho u \\ A \\ A_0
\end{pmatrix} =
\begin{pmatrix} 
  	Q_1 \\ Q_2 \\ Q_3 \\ Q_4
\end{pmatrix}, \quad
	\boldsymbol{f}(\boldsymbol{Q}) = 
\begin{pmatrix} 
 	A\rho u \\ A\rho u^2 + Ap \\ 0 \\ 0
\end{pmatrix}, \quad
	\frac{\partial \boldsymbol{f}}{\partial \boldsymbol{Q}} =
\begin{pmatrix} 
  	0 &1 &0 &0 \\ -u^2+a &2u &p+b &0 \\ 0 &0 &0 &0 \\ 0 &0 &0 &0
\end{pmatrix}, \]
	\[\boldsymbol{B}(\boldsymbol{Q}) =
\begin{pmatrix} 
  	0 &0 &0 &0 \\ 0 &0 &-p &0 \\ 0 &d &0 &0 \\ 0 &0 &0 &0
\end{pmatrix}, \quad
	\boldsymbol{A}(\boldsymbol{Q}) = 
\begin{pmatrix} 
  	0 &1 &0 &0 \\ -u^2+a &2u &b &0 \\ 0 &d &0 &0 \\ 0 &0 &0 &0
\end{pmatrix}, \]
	\[\boldsymbol{\Lambda} =
\begin{pmatrix} 
  	0 &0 &0 &0 \\ 0 &0 &0 &0 \\ 0 &0 &u - c &0 \\ 0 &0 &0 &u + c 
\end{pmatrix}, \quad
	\boldsymbol{R} =
\begin{pmatrix} 
  	-\frac{b}{a-u^2} &0 &\frac{1}{d} &\frac{1}{d} \\ 0 &0 &\frac{u-c}{d} &\frac{u+c}{d} \\ 1 &0 &1 &1 \\ 0 &1 &0 &0
\end{pmatrix}, \]
with $p = p(\rho) = p(Q_1/Q_3)$, $a = Q_3\partial p/\partial Q_1 = c_0'^2$, $b = Q_3\partial p/\partial Q_3 = -\rho c_0'^2$ and $c$ the wave speed 
\begin{equation}
\label{eq:celerity2}
	c = \sqrt{a + bd} = \frac{c_0'} {\sqrt{1+ \frac{\rho c_0'^2}{\beta A}}} = \frac{c_0'}{ \sqrt{1+ \frac{2\rho W c_0'^2}{E_0}} } ,
\end{equation}
where
\begin{equation}
\label{eq:c0'}
	c_0' = \sqrt{\frac{\partial p}{\partial \rho}} = 
	\left\{
                \begin{array}{ll}
                c_0 \quad &\text{if} \quad p \geq p_v\\
                \frac{(\rho_0 K R_v T_0 - K p)(p-p_v)-p}{\sqrt{\rho_0 K (\rho_0 T_0 R_v p_v - p^2)}} 	\quad &\text{if} \quad 0<p<p_v\\
                \end{array}
              \right.
.
\end{equation}\par
The explicit second order TVD finite volume discretisation of system (\ref{eq:sysDOT}) is:
\begin{equation}
\label{eq:FVM}
	\boldsymbol{Q}_i^{n+1} = \boldsymbol{Q}_i^{n} - \frac{\Delta t}{\Delta x}\left(\boldsymbol{f}_{i+\frac{1}{2}} - \boldsymbol{f}_{i-\frac{1}{2}}\right) - \frac{\Delta t}{\Delta x}\left(\boldsymbol{D}_{i+\frac{1}{2}} + \boldsymbol{D}_{i-\frac{1}{2}}\right) - \Delta t \boldsymbol{B}\left(\boldsymbol{Q}_i^{n+\frac{1}{2}}\right) \frac{\Delta \boldsymbol{Q}_i^n}{\Delta x} + \Delta t \boldsymbol{S}\left(\boldsymbol{Q}_i^{n+\frac{1}{2}}\right),
\end{equation}
using a uniform grid of \(N_x\) elements with mesh spacing \(\Delta x = x_{i+\frac{1}{2}}-x_{i-\frac{1}{2}}=L/N_x\) and a time step size \(\Delta t = t^{n+1}-t^{n}\) that follows the \(\CFL\) condition of Eq.~\eqref{eq:cfl}. Here the slope \( \Delta \boldsymbol{Q_i^n}\) is evaluated by using the classical minmod slope limiter \cite{Toro2009}, 
and variables at the intermediate time step \(\Delta t/2\) are calculated by the following relation:
\begin{equation}
\label{eq:Qp_t}
	\boldsymbol{Q}_i^{n+\frac{1}{2}} = \boldsymbol{Q}_i^n + \frac{1}{2} \Delta t \partial_t\boldsymbol{Q}_i^n,
\end{equation}
with the time derivative
\begin{equation}
\label{eq:dtQ}
	\partial_t \boldsymbol{Q}_i^n = - \frac{\boldsymbol{f}\left(\boldsymbol{Q}_i^n + \frac{1}{2}\Delta\boldsymbol{Q}_i^n\right) - \boldsymbol{f}\left(\boldsymbol{Q}_i^n - \frac{1}{2}\Delta\boldsymbol{Q}_i^n\right)}{\Delta x} - \boldsymbol{B}\left(\boldsymbol{Q}_i^n\right) \frac{\Delta \boldsymbol{Q}_i^n}{\Delta x} + \boldsymbol{S}\left(\boldsymbol{Q}_i^n\right);
\end{equation}
The numerical flux is obtained applying the DOT (Dumbser-Osher-Toro) solver as defined in \cite{dumbser2011a}:
\begin{equation}
\label{eq:flux}
	\boldsymbol{f}_{i+\frac{1}{2}} = \frac{1}{2} \left[ \boldsymbol{f}\left(\boldsymbol{Q}_{i+\frac{1}{2}}^{+}\right) + \boldsymbol{f}\left(\boldsymbol{Q}_{i+\frac{1}{2}}^{-}\right)\right] - \frac{1}{2} \int_{0}^{1} \left \lvert \boldsymbol{A}\left(\Psi\left(\boldsymbol{Q}_{i+\frac{1}{2}}^{-},\boldsymbol{Q}_{i+\frac{1}{2}}^{+},s\right)\right)\right \rvert \frac{\partial \Psi}{\partial s}ds,
\end{equation}
with a numerical dissipation related to matrix \(\boldsymbol{A}\) that includes both conservative and non-conservative terms.\\
The fluctuations given by the non-conservative part then read \cite{dumbser2011}:
\begin{equation}
\label{eq:fluct}
	\boldsymbol{D}_{i+\frac{1}{2}} = \frac{1}{2} \int_{0}^{1} \boldsymbol{B}\left(\Psi\left(\boldsymbol{Q}_{i+\frac{1}{2}}^{-},\boldsymbol{Q}_{i+\frac{1}{2}}^{+},s\right)\right)\frac{\partial \Psi}{\partial s}ds.
\end{equation}
The boundary-extrapolated values within cell \(i\) are given by:
\begin{equation}
\label{eq:Qbound}
	\boldsymbol{Q}_{i+\frac{1}{2}}^{-} = \boldsymbol{Q}_i^n + \frac{1}{2}\Delta	\boldsymbol{Q}_i^n + \frac{1}{2}\Delta t\partial_t\boldsymbol{Q}_i^n
	\qquad \mathrm{and} \qquad
\boldsymbol{Q}_{i-\frac{1}{2}}^{+} = \boldsymbol{Q}_i^n - \frac{1}{2}\Delta\boldsymbol{Q}_i^n + \frac{1}{2}\Delta t\partial_t\boldsymbol{Q}_i^n.
\end{equation}
The symbol \(\Psi\) stands for the path connecting left to right boundary values in the phase-space; in this work a simple linear segment has been chosen \cite{Pares2006}, hence:
\begin{equation}
\label{eq:path}
	\Psi = \Psi\left(\boldsymbol{Q}_{i+\frac{1}{2}}^{-},\boldsymbol{Q}_{i+\frac{1}{2}}^{+},s\right) = \boldsymbol{Q}_{i+\frac{1}{2}}^{-} + s\left(\boldsymbol{Q}_{i+\frac{1}{2}}^{+} - \boldsymbol{Q}_{i+\frac{1}{2}}^{-}\right).
\end{equation}
For an extension to a non-linear path scheme, the reader can refer to applications for the SWE in \cite{caleffi2017}.\\
Approximating relations (\ref{eq:flux}) and (\ref{eq:fluct}) with a Gauss-Legendre quadrature formula we get the final expressions for solving Eq.~\eqref{eq:FVM} numerically:
\begin{equation}
\label{eq:flux_num}
	\boldsymbol{f}_{i+\frac{1}{2}} = \frac{1}{2} \left[ \boldsymbol{f}\left(\boldsymbol{Q}_{i+\frac{1}{2}}^{+}\right) + \boldsymbol{f}\left(\boldsymbol{Q}_{i+\frac{1}{2}}^{-}\right)\right] - \frac{1}{2} \sum_{j=1}^{N_G}\left[\omega_j \left \lvert \boldsymbol{A}\left(\Psi\left(\boldsymbol{Q}_{i+\frac{1}{2}}^{-},\boldsymbol{Q}_{i+\frac{1}{2}}^{+},s_j\right)\right)\right \rvert \right] \left(\boldsymbol{Q}_{i+\frac{1}{2}}^{+} - \boldsymbol{Q}_{i+\frac{1}{2}}^{-}\right)
\end{equation}
and
\begin{equation}
\label{eq:fluct_num}
	\boldsymbol{D}_{i+\frac{1}{2}} = \frac{1}{2} \sum_{j=1}^{N_G}\left[\omega_j \boldsymbol{B}\left(\Psi\left(\boldsymbol{Q}_{i+\frac{1}{2}}^{-},\boldsymbol{Q}_{i+\frac{1}{2}}^{+},s_j\right)\right)\right] \left(\boldsymbol{Q}_{i+\frac{1}{2}}^{+} - \boldsymbol{Q}_{i+\frac{1}{2}}^{-}\right),
\end{equation}
where $\omega_j$ and $s_j$ are the weights and nodes of the Gauss-Legendre quadrature. In the work here presented the authors decided to use a simple 3-point  Gaussian quadrature rule (\(N_G = 3\)).

\subsection{Semi-Implicit Finite Volume Method (SI)}
\label{S:3.3}

\begin{figure}[t]
\centering
\includegraphics[width=0.35\linewidth]{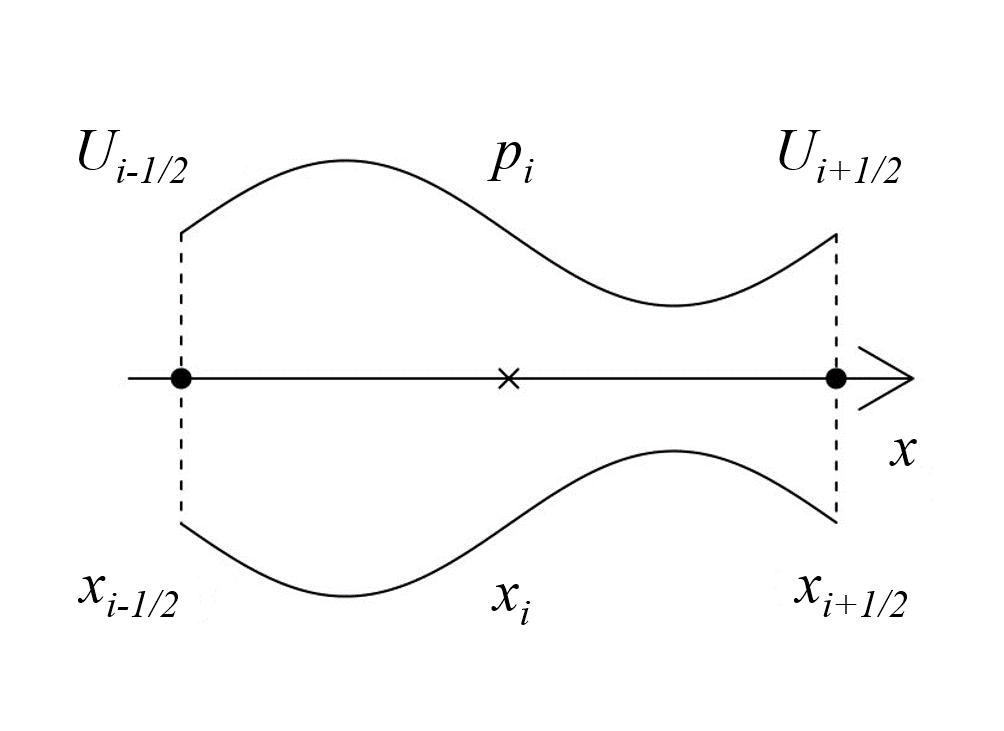}
\caption{Staggered grids for the Semi-Implicit 1D scheme: the pressure is located in the cell barycenter while the velocity is defined in the cell edges.}
\label{GridSI1D}
\end{figure}

For a Semi-Implicit numerical approach to solve the problem, we consider that along the pipe of length $L$ there are $N_x$ intervals of constant length $\Delta x=L/N_x$. The one-dimensional domain is composed by two overlapping grids according to the staggered approach: one is for the pressure and it is called main grid, and the other is for fluxes evaluation, dual mesh. The pressure is located at the cell barycenter $x_i$, meanwhile the velocities and the mass fluxes are defined at the edges $x_{i\pm\frac{1}{2}}$ of each cell (see Fig. \ref{GridSI1D}).
Then, to achieve easily second order of accuracy in time, we use the so called $\theta$-method \cite{IDIzamm}. $\theta$ is an implicitness parameter chosen in the interval $0.5\leq\theta\leq1$ for stability. In particular, when $\theta=1$ the scheme is a of first order of accuracy scheme in time and when $\theta=0.5$ the method corresponds to a Crank-Nicolson type scheme of the second order. For example, the $\theta$-method applied to the pressure $p_{i}^{n+\theta}$ gives $\theta p_{i}^{n+1} + (1-\theta)p_{i}^n$.  
The continuity equation is discretised on the main grid in a semi-implicit way: 
\begin{equation}
\rho A\left(p_i^{n+1}\right)=\rho A\left(p_i^n\right)-\frac{\Delta t}{\Delta x}\left(Q_{i+\frac{1}{2}}^{n+\theta}-Q_{i-\frac{1}{2}}^{n+\theta}\right), 
\label{cont1d}
\end{equation}
where  $\rho A\left(p_i\right)=\rho\left(p_i\right)A\left(p_i\right)$ and $Q_{i+\frac{1}{2}}^{n+1}=\rho_{i+\frac{1}{2}}^{n} A_{i+\frac{1}{2}}^{n} u_{i+\frac{1}{2}}^{n+1}$ is the
mass flow rate with $\rho_{i+\frac{1}{2}}^n=\halb \rho\left(p_i^n\right)+\halb \rho\left(p_{i+1}^n\right)$ and $A_{i+\frac{1}{2}}^n=\halb A\left(p_i^n\right)+\halb A\left(p_{i+1}^n\right)$. 
Then, the semi-implicit discretisation of momentum equation yields to:
\begin{equation}
\frac{Q_{i+\frac{1}{2}}^{n+1}-FQ_{i+\frac{1}{2}}^n}{\Delta t}=-A_{i+\frac{1}{2}}^n\frac{\Delta t}{\Delta x} \left(p_{i+1}^{n+\theta}-p_i^{n+\theta}\right)-2\pi R_{i+\frac{1}{2}}^n \rho^n_{i+\frac{1}{2}} f_{i+\frac{1}{2}}^n\frac{u_{i+\frac{1}{2}}^n u_{i+\frac{1}{2}}^{n+1}}{8} -2\pi R_{i+\frac{1}{2}}^n \left(\tau_u^n\right)_{i+\frac{1}{2}}, 
\label{Mom1DSI}
\end{equation}
where $FQ_{i+\frac{1}{2}}^n$ is an explicit and nonlinear operator for the convective terms. Here we consider a robust explicit upwind approach based on the Rusanov method which allows also to keep the well-balancing properties of the flux as done in \cite{dumbser2015,IDIzamm}:
\begin{equation}
\begin{split}
& FQ_{i+\halb}^n=Q_{i+\halb}^n-\frac{\Delta t}{\Delta x}\left(f_{i+1}^{Rus,n}-f_{i}^{Rus,n}\right), \quad \text{with}\\
& f_{i}^{Rus,n}=\frac{1}{2}\left(u_{i+\halb}^n Q_{i+\halb}^n+u_{i-\halb}^n Q_{i-\halb}^n\right)-\frac{1}{2}S_{\max}\left(Q_{i+\halb}^n-Q_{i-\halb}^n\right), \qquad S_{\max}=2 \max\left(|u_{i-\halb}^n|,|u_{i+\halb}^n|\right). \\
\end{split}
\label{eqn.ruf} 
\end{equation}
In addition we rewrite Eq.~\eqref{Mom1DSI} as  
\begin{equation}
Q_{i+\frac{1}{2}}^{n+1}=G_{i+\frac{1}{2}}^{n}- \theta A_{i+\frac{1}{2}}^n \frac{\Delta t}{\Delta x}\left(p_{i+1}^{n+1}-p_{i}^{n+1}\right) -\Delta t \gamma_{i+\frac{1}{2}}^n Q_{i+\frac{1}{2}}^{n+1}, 
\label{mom1Dsi}
\end{equation}
where the term $\gamma_{i+\frac{1}{2}}^n=\frac{2 \pi R_{i+\frac{1}{2}}^n f_{i+\frac{1}{2}}^n |u_{i+\frac{1}{2}}^n|}{8A_{i+\frac{1}{2}}^n} \geq 0$ accounts the explicit contribution of the quasi-steady friction and $G_{i+\frac{1}{2}}^n$ collects all the explicit terms:
\begin{equation}
G_{i+\frac{1}{2}}^n={FQ}_{i+\frac{1}{2}}^n - \left(1-\theta\right) A_{i+\frac{1}{2}}^n \frac{\Delta t}{\Delta x}\left(p_{i+1}^{n}-p_{i}^{n}\right)-2\pi R_{i+\frac{1}{2}}^n \left(\tau_{u}\right)^n_{i+\frac{1}{2}}\Delta t. 
\label{G1D}
\end{equation}
The unsteady friction terms $\tau_u^n$ is computed with the approximation of the Zielke integral presented in section \ref{S:2}. 
Later on, collecting all the quantities with $Q_{i+\frac{1}{2}}^{n+1}$ at the left hand side yields the following expression:
\begin{equation}
Q_{i+\frac{1}{2}}^{n+1}=\left(   \frac{G}{1+\Delta t \gamma}  \right)_{i+\frac{1}{2}}^n - \theta \frac{\Delta t}{\Delta x} \left( \frac{A}{1+\Delta t \gamma} \right)_{i+\frac{1}{2}}^n \left(p_{i+1}^{n+1} -p_{i}^{n+1}\right).
\label{q1d}
\end{equation} 
Coupling Eq. \eqref{q1d} and Eq. \eqref{cont1d} gives
\begin{equation}
\rho A\left(p_{i}^{n+1}\right)-\theta^2 \frac{{\Delta t}^2}{\Delta x^2}\left[ \left(p_{i+1}^{n+1}-p_{i}^{n+1}\right)\left(   \frac{
A}{1+\Delta t \gamma}  \right)_{i+\frac{1}{2}}^n    -\left(p_{i}^{n+1}-p_{i-1}^{n+1}\right)\left(   \frac{
A}{1+\Delta t \gamma}  \right)_{i-\frac{1}{2}}^n   \right] =b_i^n, 
\label{contdiscr1D}
\end{equation}
which can be written as a mildly non-linear system of equations 
\begin{equation}
\rho \boldsymbol{A}\left(\boldsymbol{p}^{n+1}\right) + \boldsymbol{T} \boldsymbol{p}^{n+1}=\boldsymbol{b}\left(\boldsymbol{p}^n\right),  
\label{MNL}
\end{equation}
where $\rho \boldsymbol{A}$ is the non-linear diagonal contribution, $\boldsymbol{T}$ is the linear and symmetric three-diagonal part, $\boldsymbol{p}^{n+1}$ is unknown vector pressure
and $\boldsymbol{b}^n$ is the known right hand side term. System \eqref{MNL} can be solved by using a Newton-type algorithm such as the one of Brugnano and Casulli \cite{BrugnanoCasulli2008,BrugnanoCasulli2009} or the more general one of Casulli and Zanolli \cite{CasulliZanolli2010,CasulliZanolli2012}. For more details see \cite{IDIzamm,dumbser2015,TDC2013,FDC2014,CDT2012}.
The density is updated using the closures in Eq.~\eqref{eq:eos} while the cross sectional area is updated using the Laplace law \eqref{eq:laplace} or the ODE versions of Eq.~\eqref{eq:3par.ODE} and Eq.~\eqref{eq:npar.ODE} respectively  discretised as follows:
\begin{equation}
A_i^{n+1}=A_i^{n}+\frac{2WA_0}{E_0}\left(p_i^{n+1}-p_i^{n}\right) +\Delta t \left[ \frac{ 2WA_0p_i^{n+1}}{\tau_r E_0}-\frac{(A_i^n-A_0)E_\infty}{\tau_r E_0}\right]
\end{equation}
\begin{equation}
A_i^{n+1}=A_i^{n}+\frac{2WA_0}{E_0} \left(p_i^{n+1}-p_i^{n}\right) + 2A_0\Delta t\left[p_i^{n+1}W \sum_{k=1}^{N_{KV}}\frac{1}{\eta_k} - \sum_{k=1}^{N_{KV}}\frac{\epsilon_{rk}\left(p_i^{n+1},p_i^{n}\right)}{\tau_{rk}} \right].
\end{equation}
Finally, the velocity is computed as:
\[u_{i+\frac{1}{2}}^{n+1}=\frac{Q_{i+\frac{1}{2}}^{n+1}}{\rho_{i+\frac{1}{2}}^n A_{i+\frac{1}{2}}^n}. \]
The time step $\Delta t$ for the Semi-Implicit model is the one given by the stability condition for computation of the non-linear convective terms. In this case we have that $\Delta t =\CFL \frac{\Delta x}{2|u_{max}|} $, instead of the standard \(\CFL\) condition defined with Eq.~\eqref{eq:cfl}. We emphasize that this condition is based only on the fluid velocity and not on the speed of sound which makes this method very efficient especially for the low Mach number regime. In addition, for some simulations the contribution of the convection can be neglected, $FQ=Q$, and the scheme becomes unconditionally stable. However, the time step $\Delta t$ has to be chosen properly to reduce the numerical viscosity of the method.  


\section{Unsteady friction effects and ODE Model validation}
\label{S:33}
\begin{figure}[t!]
\centering
\includegraphics[width=0.6\linewidth]{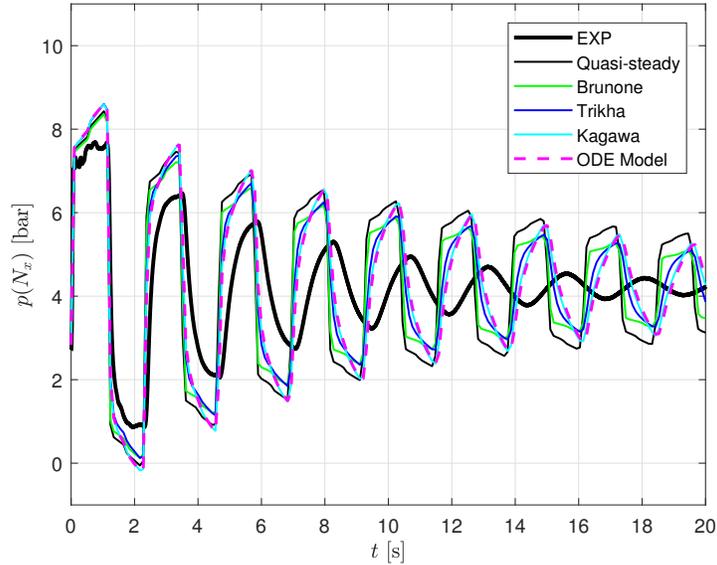}
\caption{Experimental data compared against numerical results obtained with different friction models for transients in HDPE DN50 smooth-wall pipe with turbulent flow (\(Q_0\)~=~2,00~l/s, \(\Re \approx\)~51000). Pressure \(p(N_x)\) at the downstream end vs time. The reader is advised to refer to the coloured figures of the electronic version of this paper.}
\label{ODE-US}
\end{figure}

In order to better analyse the effects of the unsteady friction term and to validate the ODE Model presented in section \ref{S.S:FR} for turbulent flow cases, we consider a water hammer test case assuming as first attempt that the pressure damping is only determined by the friction losses, neglecting the visco-elastic effects. In Fig.~\ref{ODE-US} the classical water hammer solution obtained considering only the quasi-steady friction term in Eq.~\eqref{eq:tau} is presented together with the solutions derived taking into account the complete expression of the equation, using different unsteady friction models. The reference solutions taken into account are represented by Brunone's model, Thrika's and Kagawa's formulation. In the same figure, the experimental curve is also represented. \\
The simulations are run using only the Explicit Path-Conservative Method presented in section \ref{S:3.2}, since the unsteady friction models behave in the same way in all the numerical schemes, not being affected by the chosen numerical discretisation. \par
Brunone's Model \cite{Brunone2000} is part of the Instantaneous Acceleration (IA) methods, based on the hypothesis that the unsteady wall shear stress is directly proportional to the acceleration of the flow, thus:
\begin{equation}
\label{US:Brunone}
\tau_u = \frac{\rho DK_{Bru}}{4}\left[ \frac{\partial u}{\partial t} + sign\left( u \frac{\partial u}{\partial x}\right) c_0 \frac{\partial u}{\partial x}\right],
\end{equation}
considering for the coefficient \(K_{Bru}\) the expression suggested by Vardy and Brown:
\begin{equation}
K_{Bru} = 0.5 \sqrt{\frac{7.41}{\Re^{\chi}}}, \quad \quad \chi = \log\left(\frac{14.3}{\Re^{0.05}}\right).
\end{equation}
Trikha's and Kagawa's formulations belong to the class of the Convolution Integral (CI) methods, as the ODE Model presented in this paper, for which the analytic expression for the calculation of the unsteady losses is given by the convolution integral of Zielke, Eq.~\eqref{eq:tau_zielke}. To solve this integral in turbulent flow conditions, first Trikha proposed to use the same approach adopted for the laminar case with the following approximated weighting function \cite{trikha1975}:
\begin{equation}
w(t) = \sum_{i=1}^{N_w}m_i \exp\left(-n_i\frac{\nu t}{R^2}\right)
\end{equation}
with \(N_w\) = 3, (\(m_1, m_2, m_3\)) = (40.0, 8.1, 1) and (\(n_1, n_2, n_3\)) = (8000, 200, 26.4). Writing the weighting function as a series of exponential functions, the unsteady wall shear stress calculated with Trikha's formulation becomes:
\begin{equation}
\tau_u^{n+1} \approx \frac{2\mu}{R} \sum_{i=1}^3 \tau_i^{n+1} = \frac{2\mu}{R} \left[ \sum_{i=1}^3 \exp \left(-n_i \frac{\nu t}{R^2}\right) \tau_i^n + \sum_{i=1}^3 m_i\left(u^{n+1} -u^n\right) \right] .
\end{equation}
Successively, Kagawa proposed a more efficient formulation for approximating the convolution integral \cite{Kagawa}. Considering the weighting function for turbulent case presented in section \ref{S.S:FR} in Eq.~\eqref{wapp} with Urbanowicz and Zarzycki coefficients, Kagawa's solution becomes:
\begin{equation}
\tau_u^{n+1} \approx \frac{2\mu}{R} \sum_{i=1}^{N_w} \tau_i^{n+1} = \frac{2\mu}{R} \left[ \sum_{i=1}^{N_w}  \exp \left(-(n_i^* + B^*) \frac{\nu t}{R^2}\right) \tau_i^n + \sum_{i=1}^{N_w} A^*m_i^* \exp \left(-(n_i^* + B^*) \frac{\nu t}{R^2}\right) \left(u^{n+1} -u^n\right) \right] .
\end{equation}
\par
It can easily be observed that with none of the friction formulations it is possible to completely describe the dampening behaviour of an HDPE pipe in case of hydraulic transients, confirming that the visco-elastic effects must be taken into account to obtain realistic solutions. The shape of the pressure wave appears significantly different than the exact solution especially with Brunone's model, while the major damping is given using Trikha's formulation. \\
Nevertheless, with this test, we can confirm that the ODE Model reproduces reliable results if compared to the other unsteady friction models. As expected, adopting the ODE Model we obtain the same solution given by Kagawa's formulation. Both these two unsteady friction models, indeed, belong to the CI methods category and are based on the same weighting function and coefficients. It is worth remembering that the advantage of choosing the ODE Model is explained in terms of computational cost, as already discussed in \cite{IDIzamm} for laminar flow cases.

\section{Calibration of the visco-elastic parameters}
\label{S:4}
For the water hammer test cases, calibration of the visco-elastic parameters is necessary to accurately reproduce the behaviour of the pipe material. The instantaneous elastic modulus \(E_0\) is estimated accordingly to the reference elastic wave speed value of each test. As a matter of fact, knowing the mean value of the wave speed, estimated by observing the oscillation period on the basis of experimental measurements, and using the definition \eqref{eq:celerity2}, it is possible to obtain the proper value of \(E_0\) \cite{evangelista2015}. Concerning the rest of the visco-elastic parameters, for a multi-parameter model, while \(\tau_k\) are fixed as in references \cite{covas2005,evangelista2015}, \(E_k\) are calibrated by minimizing the least square error (LSE) between numerical and experimental pressure at the downstream end. The same principle is followed for the calibration of \(E_{\infty}\) and \(\eta\) with the 3-parameter model. To carry out these optimizations the SCE-UA (Shuffled Complex Evolution~-~University of Arizona) algorithm, a general purpose global optimization method originally developed by Duan et al. \cite{duan1992,duan1993} has been used.\par
Two main approaches were followed to calibrate the creep function and test the numerical models. Having observed in section \ref{S:33} that in general the unsteady friction term cannot extensively describe the dissipation of transient waves in HDPE pipes, in the first calibration we neglected the unsteady friction effects, considering only the pipe wall visco-elasticity as diffusive effect. In the second calibration, instead, we considered the unsteady friction losses as part of the dampening. It has been noticed that the calibration of the visco-elastic parameters is not independent of the specific test facilities, in terms of diameter and length of the tube, wall thickness and anchors. Hence, in order to achieve the best fitting between numerical and experimental results, a specific calibration has been made for each test analysed, considering not to have parameters generally valid for a given material. Moreover, it is worth to mention that the existence of different combinations of visco-elastic parameters for describing the behaviour of a plastic tube, with the same sum of squared errors against experimental data, has recently been confirmed by Ferrante and Capponi \cite{ferrante2017} when using the SLSM.\\
Visco-elastic calibrated parameters are further presented in section \ref{S:5.1} for each water hammer test case.


\section{Numerical results}
\label{S:5}
To compare the three numerical methods, two different types of test problems have been selected.\\ The first kind of test cases regard three different Riemann problems, solved only for the elastic case, hence using the Laplace law, for which a quasi-exact solution is available \cite{dumbser2015}.\\
Secondly, two water hammer problems in HDPE tubes are presented, for which experimental data already used by Pignatelli \cite{Pignatelli2014} and Evangelista et al. \cite{evangelista2015} were provided and assumed as reference. For this kind of tests, visco-elastic parameters calibration is also discussed.\par
In all the simulations presented in this paper the following assumptions are considered: CFL = 0.9, \(\nu_p\)~=~0.4, \(\rho_0\) = 998.2 kg/m$^3$, \(p_0\) = \(10^5\) Pa, \(c_0\) = 1400 m/s, \(T_0\) = 293 K, \(p_v\) = 2300 Pa, \(R_v\) = 303 JK$^{-1}$/kg and \(K\) = \(10^{-6}\) Pa $^{-1}$.


\begin{figure}[t!]
\begin{subfigure}{0.45\textwidth}
\centering
\includegraphics[width=1\linewidth]{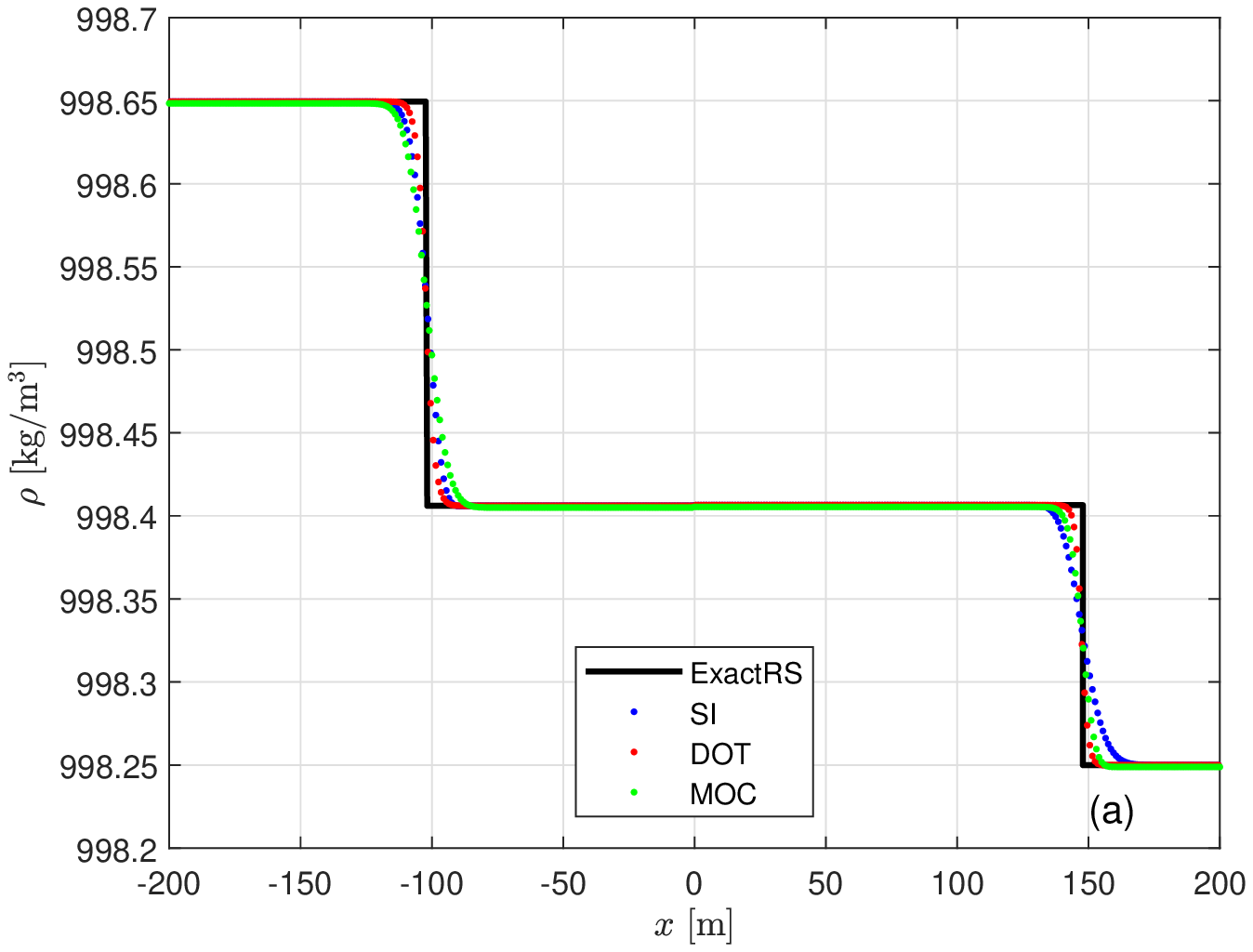}
\label{fig.RP1rho}
\end{subfigure}
\hspace{7mm}
\begin{subfigure}{0.45\textwidth}
\centering
\includegraphics[width=1\linewidth]{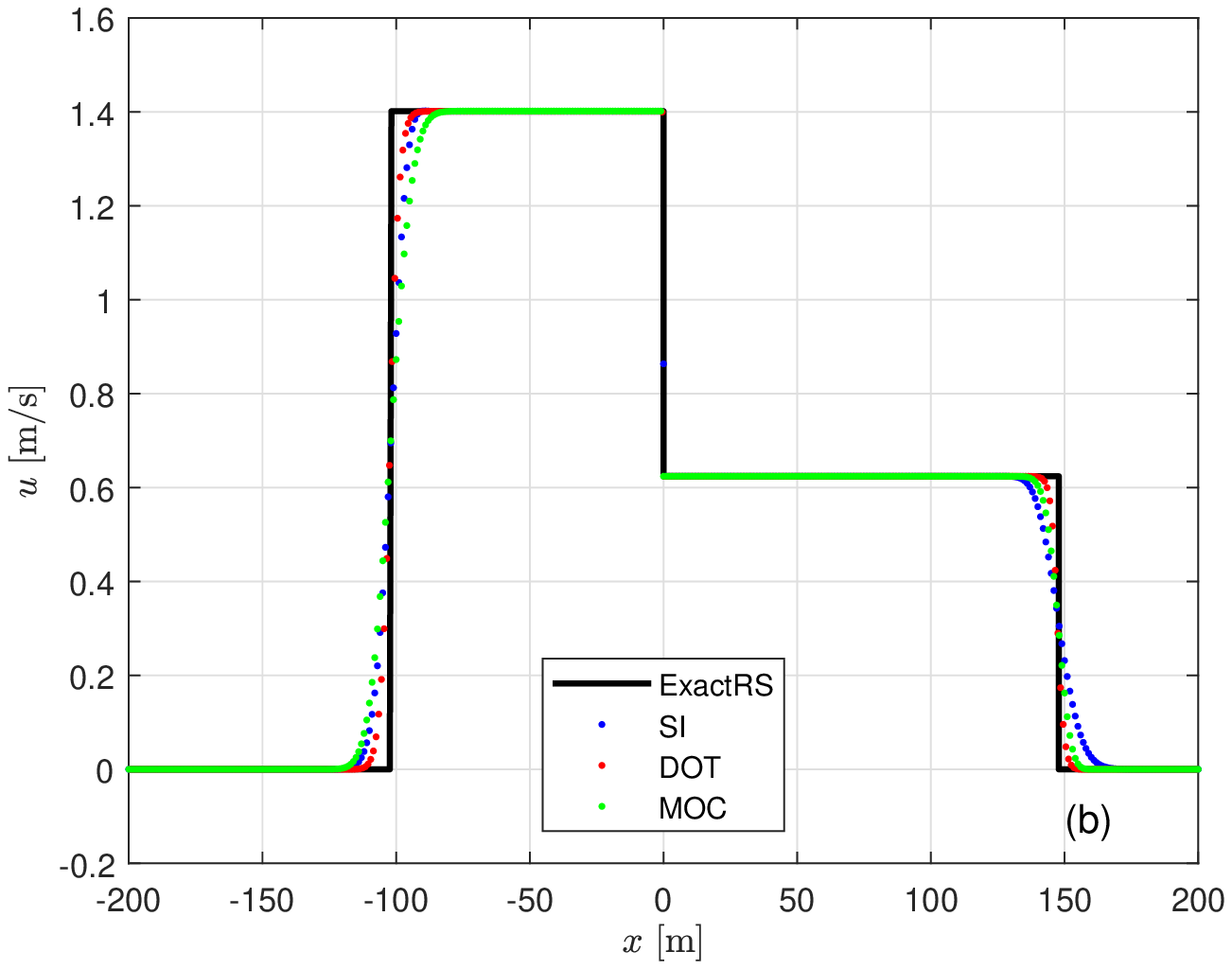}
\label{fig.RP1u}
\end{subfigure}
\hfill
\vspace{6mm}
\begin{subfigure}{0.45\textwidth}
\centering
\includegraphics[width=1\linewidth]{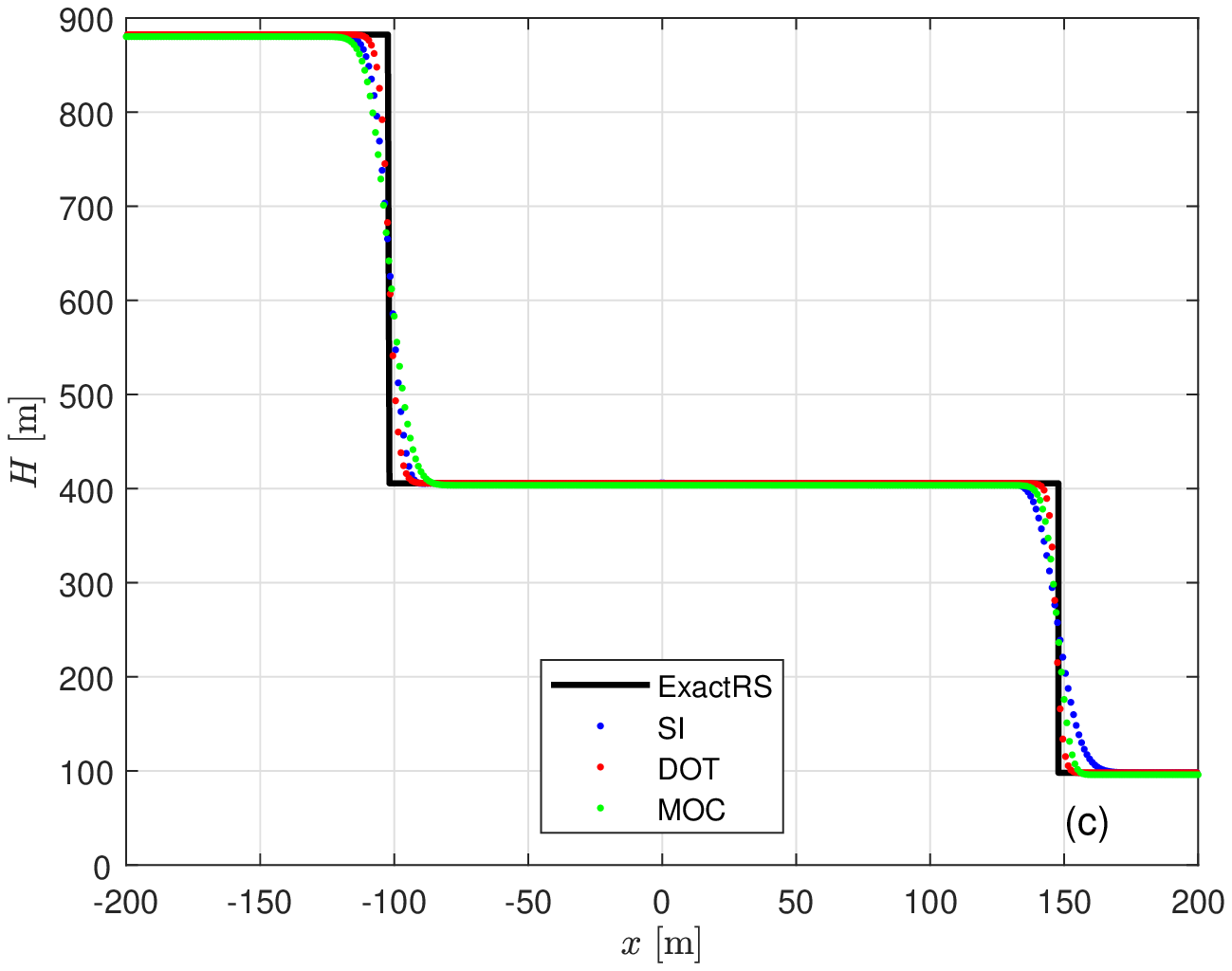}
\label{fig.RP1TH}
\end{subfigure}
\hspace{7mm}
\begin{subfigure}{0.45\textwidth}
\centering
\includegraphics[width=1\linewidth]{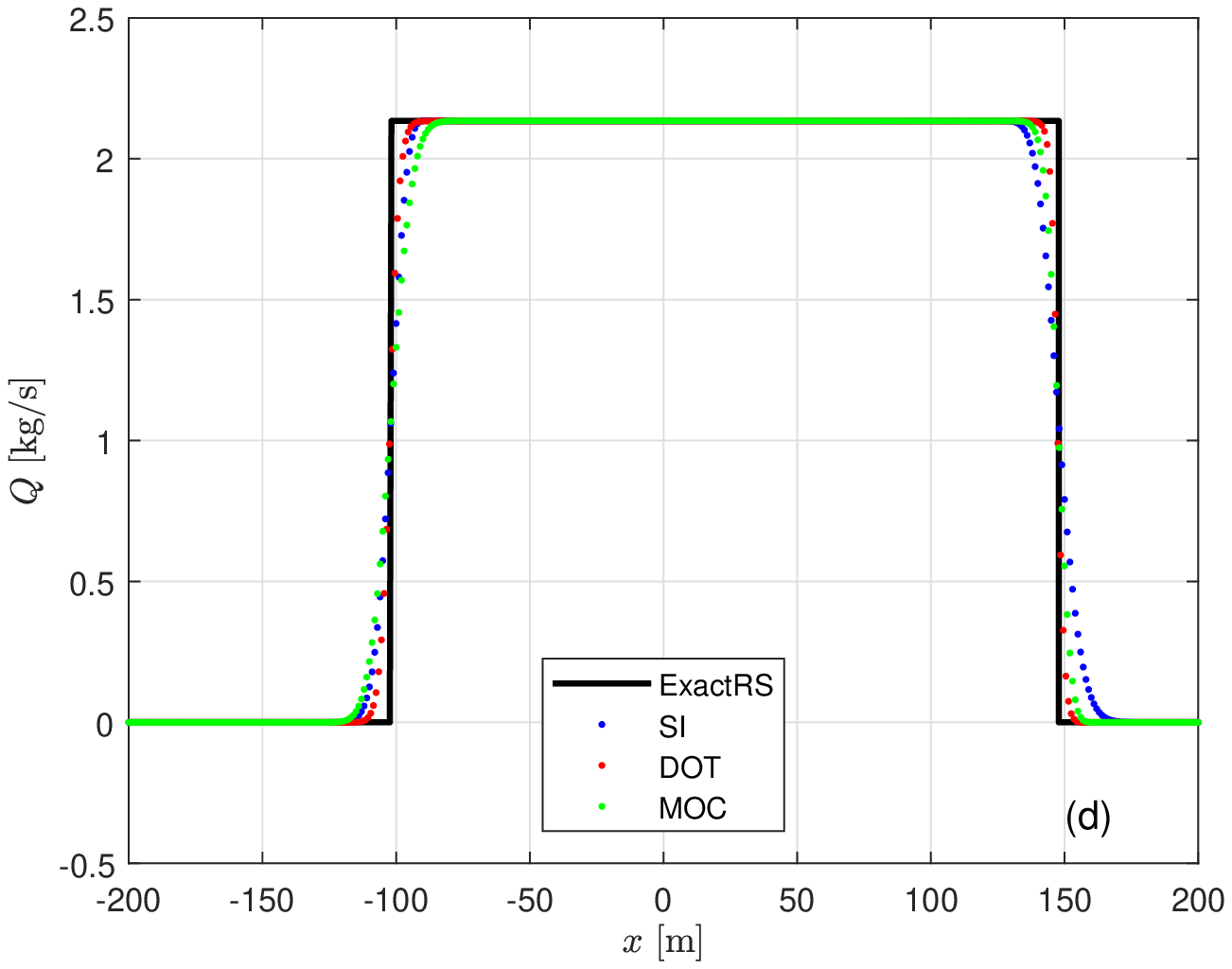}
\label{fig.RP1Q}
\end{subfigure}
\caption{Comparison of the numerical results obtained with MOC, DOT and SI against the quasi-exact solution (ExactRS) in the Riemann problem RP1 at time \(t_{end}\) in terms of (a) density, (b) velocity, (c) total head and (d) flow rate. The reader is advised to refer to the coloured figures of the electronic version of this paper.}
\label{fig.RP1}
\end{figure}

\begin{figure}[t!]
\begin{subfigure}{0.45\textwidth}
\centering
\includegraphics[width=1\linewidth]{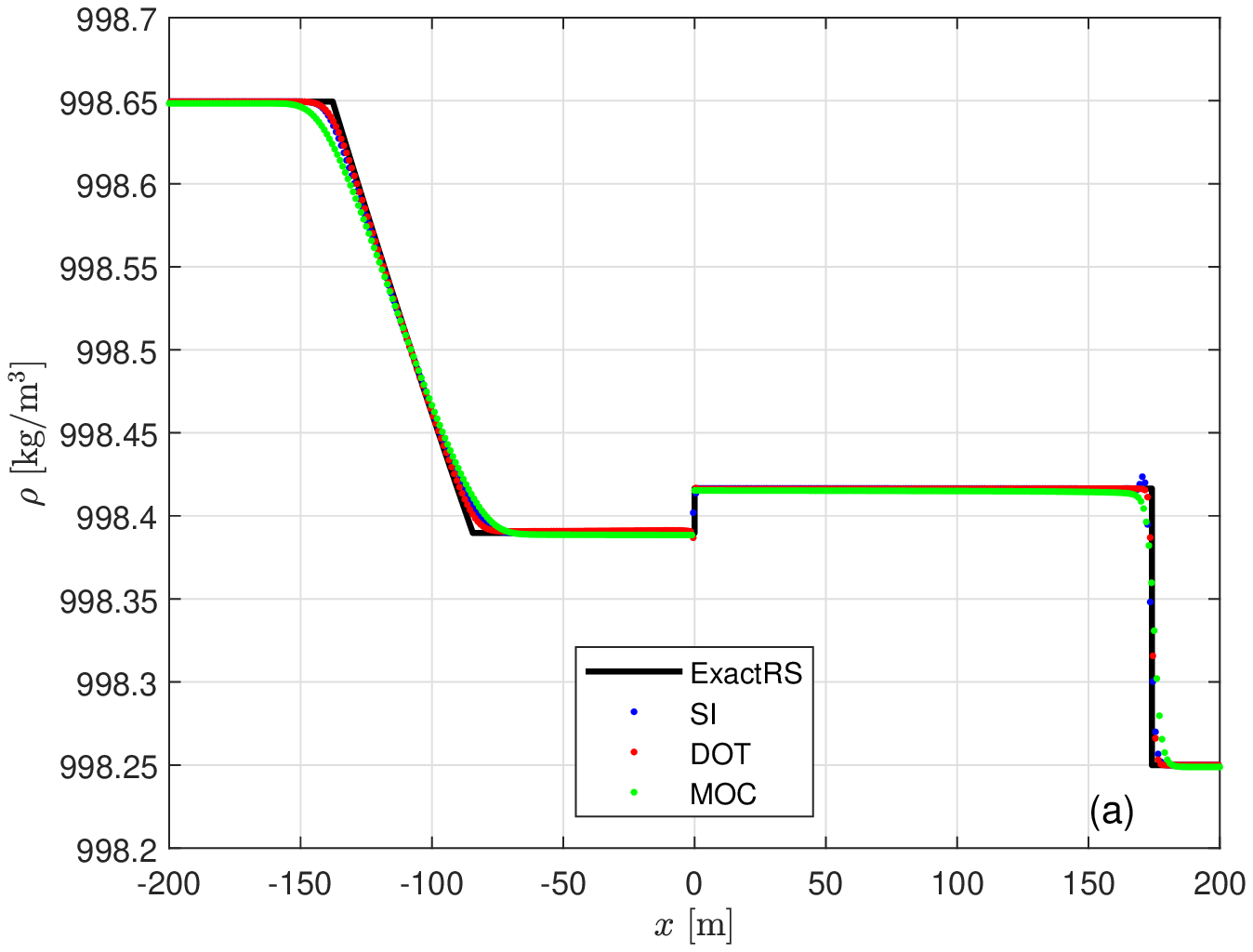}
\label{fig.RP2rho}
\end{subfigure}
\hspace{7mm}
\begin{subfigure}{0.45\textwidth}
\centering
\includegraphics[width=1\linewidth]{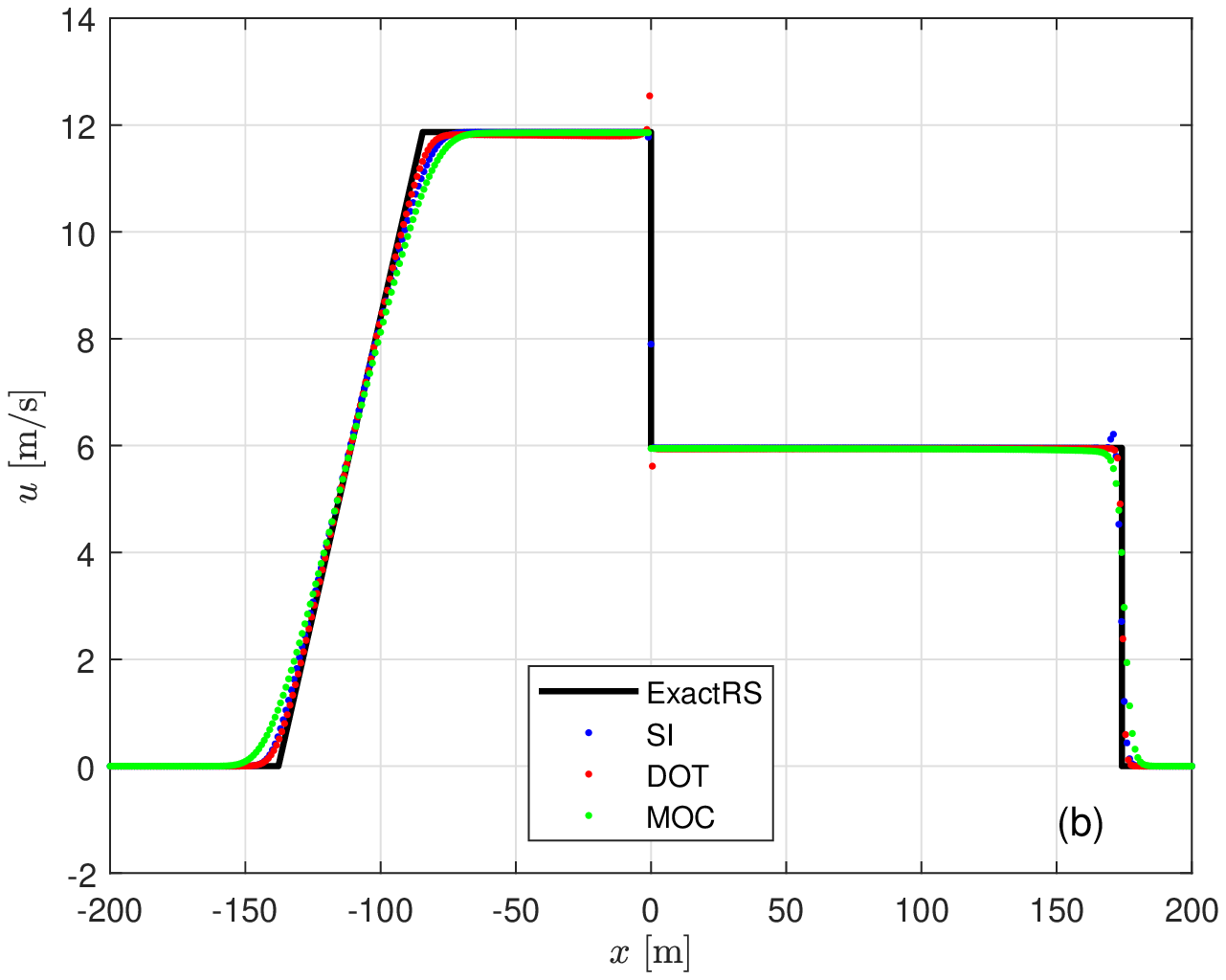}
\label{fig.RP2u}
\end{subfigure}
\hfill
\vspace{6mm}
\begin{subfigure}{0.45\textwidth}
\centering
\includegraphics[width=1\linewidth]{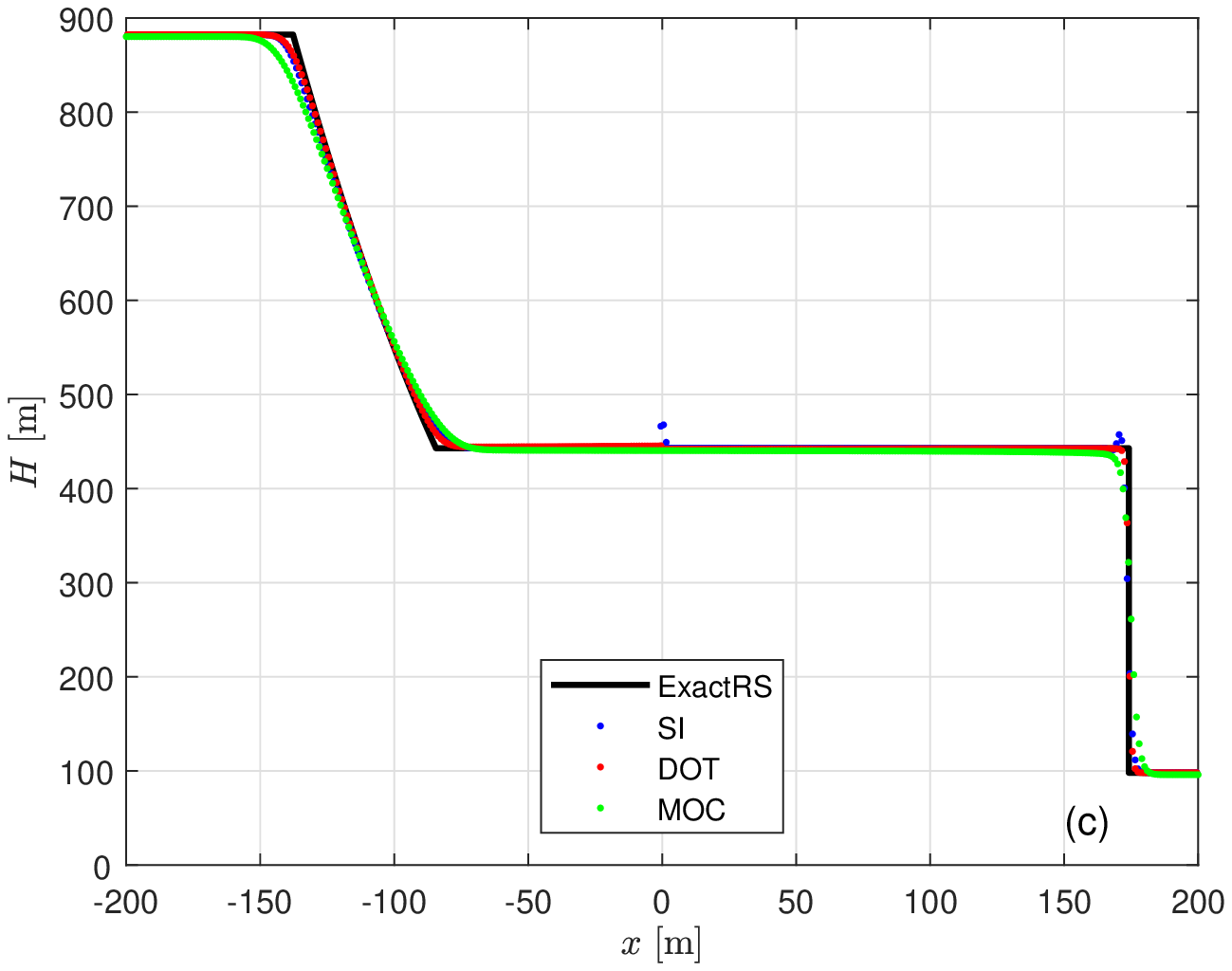}
\label{fig.RP2TH}
\end{subfigure}
\hspace{7mm}
\begin{subfigure}{0.45\textwidth}
\centering
\includegraphics[width=1\linewidth]{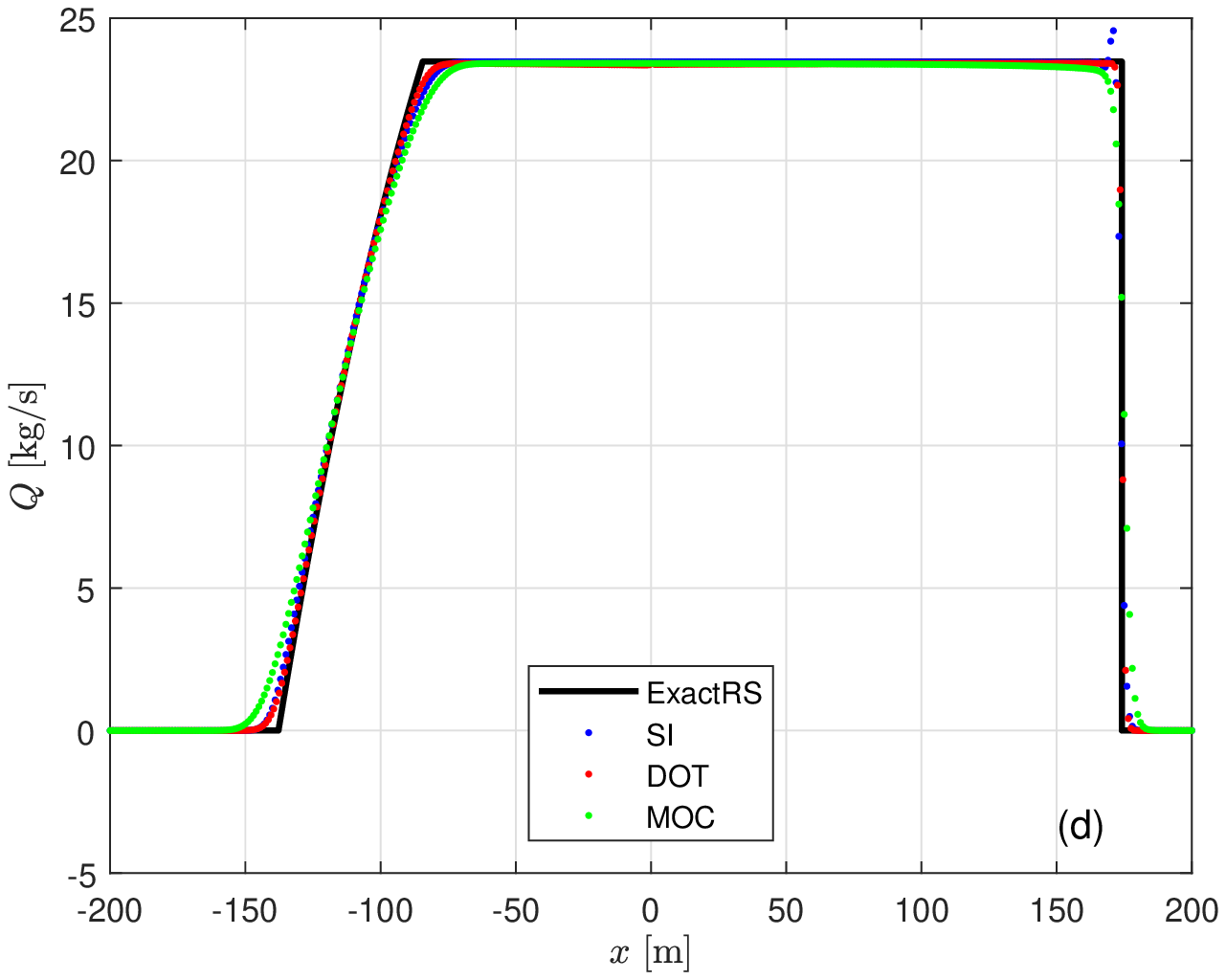}
\label{fig.RP2Q}
\end{subfigure}
\caption{Comparison of the numerical results obtained with MOC, DOT and SI against the quasi-exact solution (ExactRS) in the Riemann problem RP2 at time \(t_{end}\) in terms of (a) density, (b) velocity, (c) total head and (d) flow rate. The reader is advised to refer to the coloured figures of the electronic version of this paper.}
\label{fig.RP2}
\end{figure}

\begin{figure}[t!]
\begin{subfigure}{0.45\textwidth}
\centering
\includegraphics[width=1\linewidth]{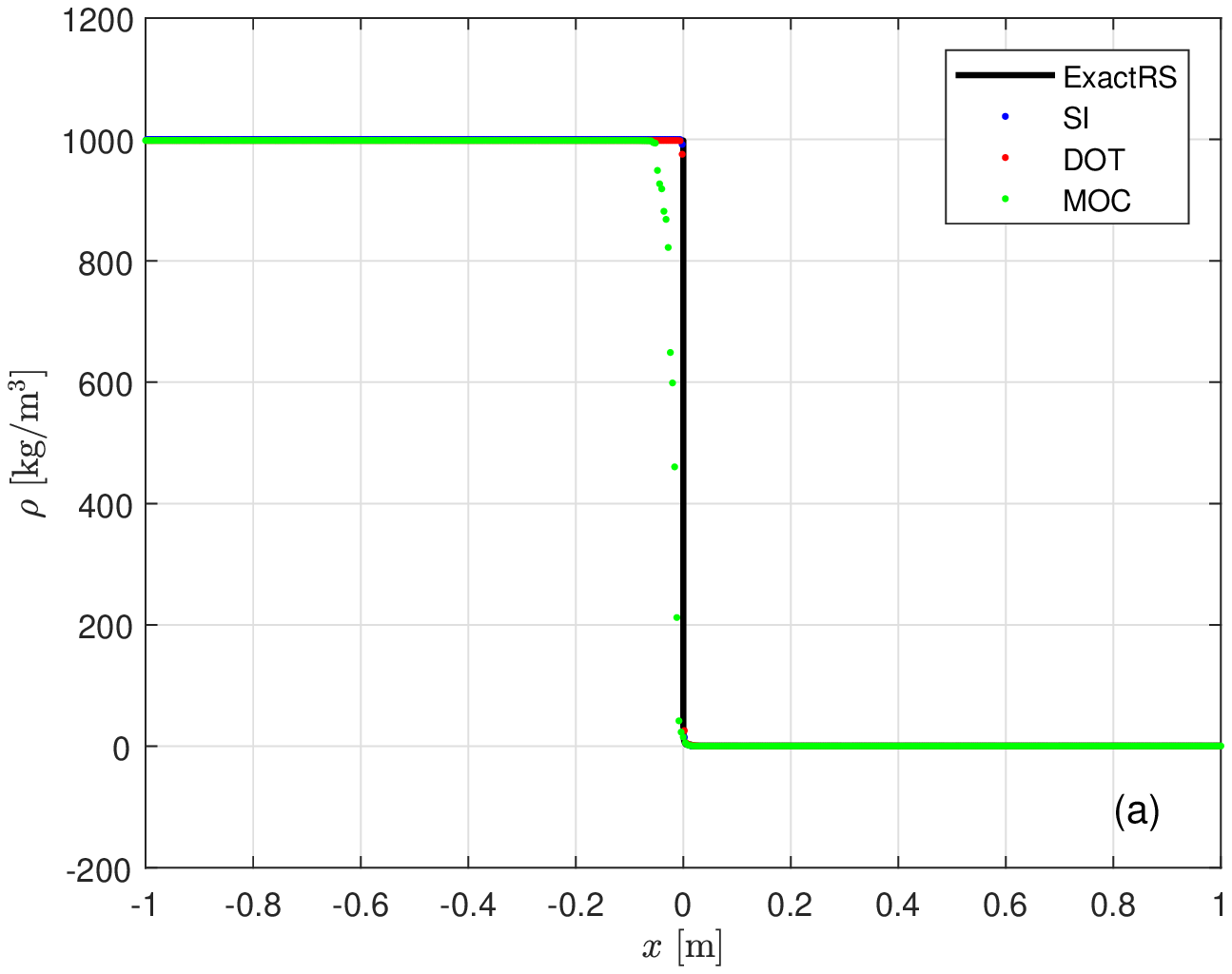}
\label{fig.RP3rho}
\end{subfigure}
\hspace{7mm}
\begin{subfigure}{0.45\textwidth}
\centering
\includegraphics[width=1\linewidth]{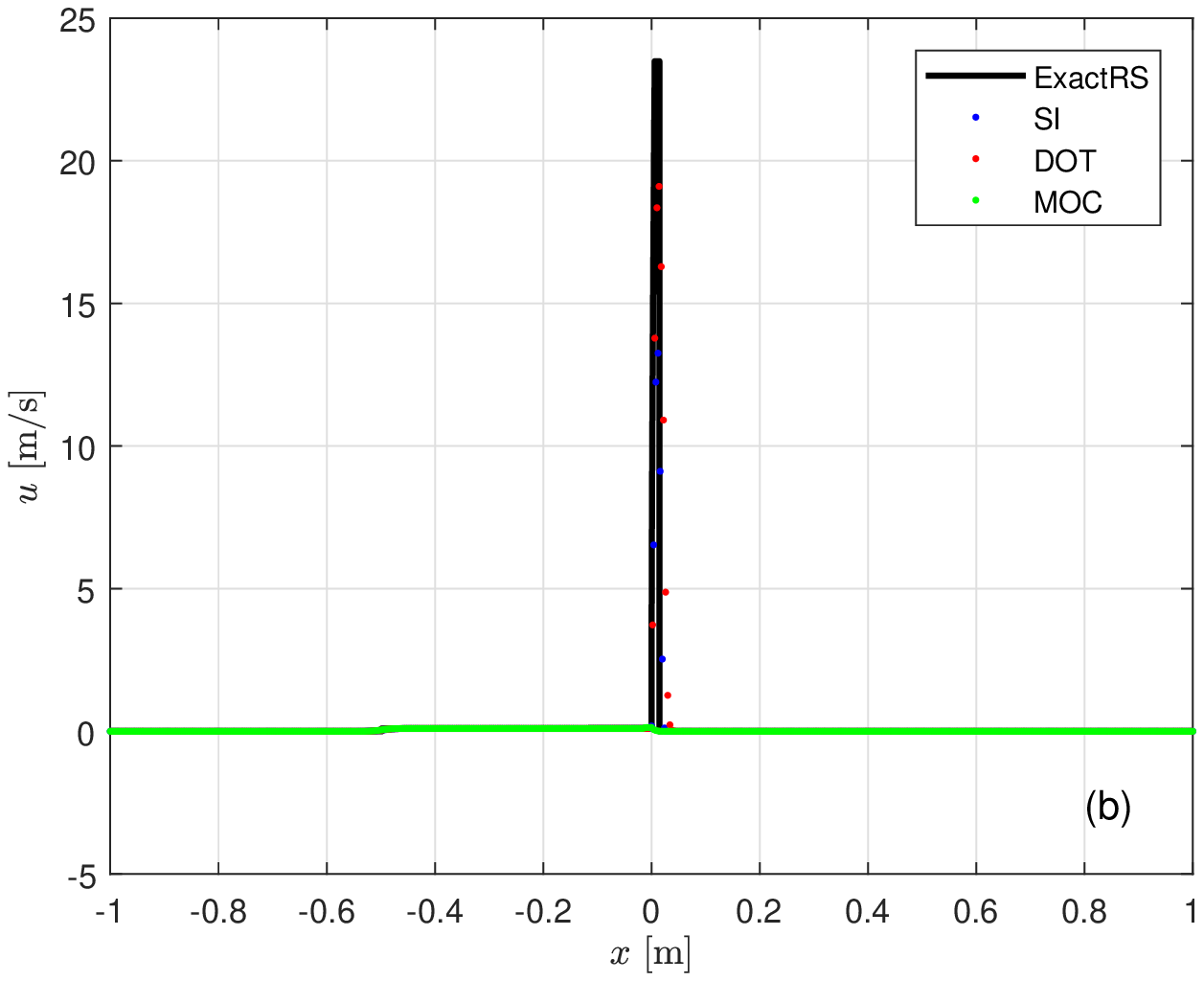}
\label{fig.RP3u}
\end{subfigure}
\hfill
\vspace{0.6cm}
\begin{subfigure}{0.45\textwidth}
\centering
\includegraphics[width=1\linewidth]{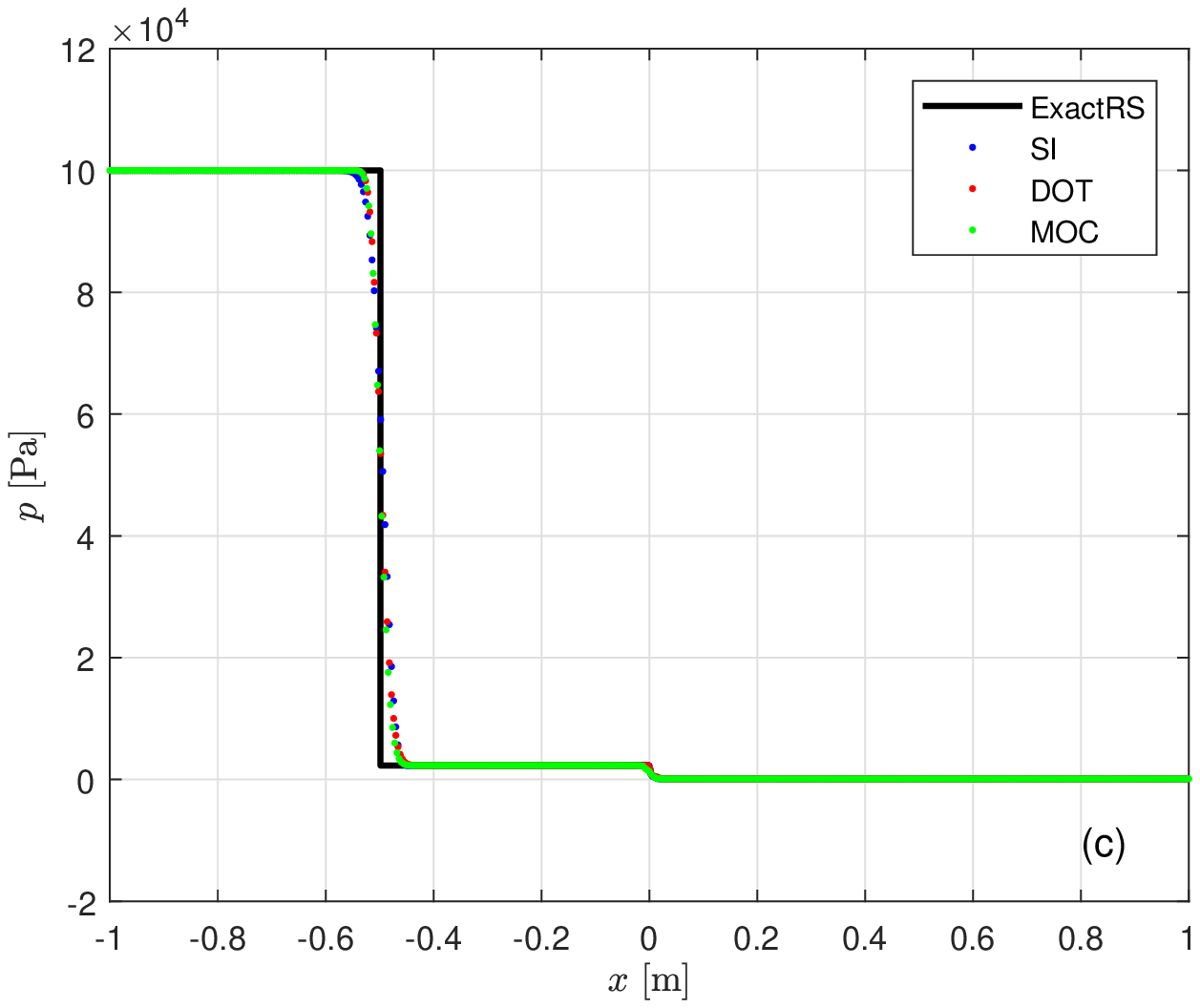}
\label{fig.RP3p}
\end{subfigure}
\hspace{7mm}
\begin{subfigure}{0.45\textwidth}
\centering
\includegraphics[width=1\linewidth]{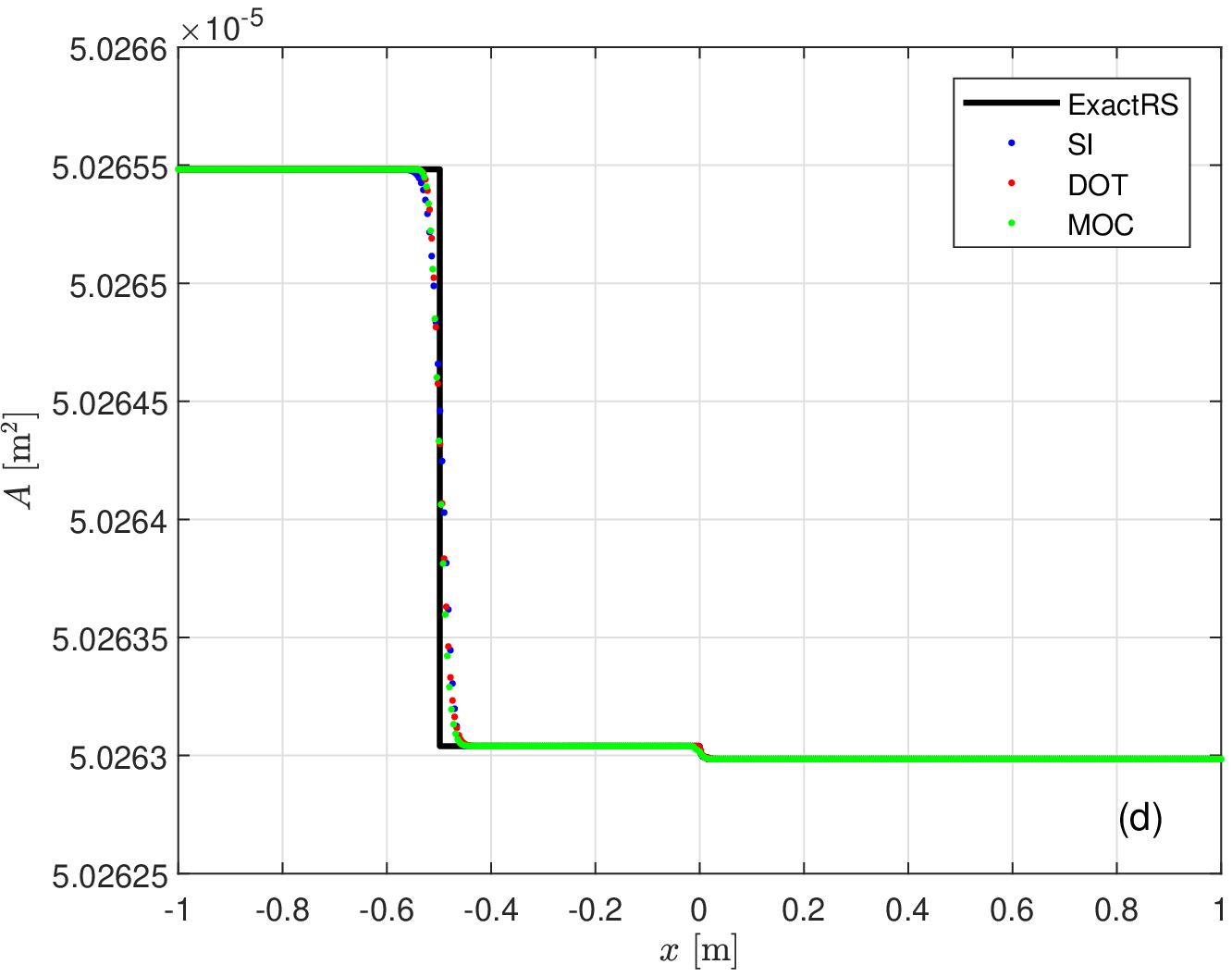}
\label{fig.RP3A}
\end{subfigure}
\caption{Comparison of the numerical results obtained with MOC, DOT and SI against the quasi-exact solution (ExactRS) in the Riemann problem RP3 at time \(t_{end}\) in terms of (a) density, (b) velocity, (c) pressure and (d) cross-sectional area. The reader is advised to refer to the coloured figures of the electronic version of this paper.}
\label{fig.RP3}
\end{figure}

\begin{table}[b!]
\centering
\begin{tabular}{l P{1cm} P{1cm} P{2cm} P{1cm} P{1cm} P{2cm} P{1.2cm} P{1cm} P{1cm}}
\hline
	 Case &\(p_L\) &\(u_L\) &\(A_{0L}\) &\(p_R\) &\(u_R\) &\(A_{0R}\) &\(\beta\) &\(t_{end}\) &\(N_x\)\\
\hline
	\textbf{RP1} &100\(\cdot \gamma_0\) & 0.0 & 0.0015 & 20\(\cdot \gamma_0\) & 0.0 & 0.0034 & \(8\cdot 10^{10}\) &0.3 &400 \\
	\textbf{RP2} &100\(\cdot \gamma_0\) & 0.0 & 0.0015 & 20\(\cdot \gamma_0\) & 0.0 & 0.0034 & \(8\cdot 10^{8}\) &3.0 &400 \\
	\textbf{RP3} &\(10^5\) & 0.0 & \(5.0265 \cdot 10^{-5}\) &\(10^2\) & 0.0 & \(5.0265\cdot 10^{-5}\) & \(4\cdot 10^{13}\) &5.0\(\cdot 10^{-4}\) & 500 \\
\hline
\end{tabular}
\caption{Initial states for the Riemann problems, with \(\gamma_0 = \rho_0 g\). Subscripts \(L\) and \(R\) stand respectively for left and right state of the piece-wise constant initial values typical of Riemann problems. Units of measurement considered: $p$ [Pa], $u$ [m/s], $A_0$ [m$^2$], $\beta$ [Pa/m$^2$], $t_{end}$ [s].}
\label{tab.RPdata}
\end{table}

\subsection{Riemann problems}
\label{S:5.2}

The chosen Riemann problems are very demanding test cases and have been run to stress the numerical schemes and evaluate their possible weaknesses.\\
The first two Riemann problems, RP1 and RP2, are set up considering a sudden increment of the cross-section of the conduct in the middle of the domain and differs each other only for the material of the pipe taken into account: RP1 concerns an elastic modulus typical of polymer tubes, while in RP2 we consider a more flexible rubber duct. The solution of the problem consists in a rarefaction wave followed by a central contact discontinuity in the middle of the domain (due to the cross-sectional jump) and a final shock wave. The differences in the two solutions are connected to the different material properties: only concerning a very flexible material the rarefaction wave appears well extended, while in the first case the rarefaction could be confused for a shock wave. \\
The last Riemann problem, RP3, is related to a general elastic flexible pipe in which cavitation occurs and hence the fluid consists of a mixture of liquid and vapour. In this problem we can observe a very strong rarefaction, which travels also through the phase change, followed by an equally severe shock wave (especially visible in the velocity plot).\\
All of these tests are solved only considering the Laplace constitutive law, to make possible the comparison of the results with a quasi-exact solution \cite{dumbser2015}. Initial data of each Riemann problem are presented in Tab.~\ref{tab.RPdata} and the final results are shown in Figs.~\ref{fig.RP1}-\ref{fig.RP3}. For the SI scheme, \(\theta = 0.80\) in all the Riemann problem simulations, \(\Delta t_{max} = 0.001\)~s in RP1 and RP2, while \(\Delta t_{max} = 0.000001\)~s in RP3. It has to be mentioned that in order to obtain a reliable result solving RP2 and RP3 with the Method of Characteristics, it was necessary to modify the code considering non-straight characteristic curves, and thus \(dx/dt = u\pm c\). This aspect has to be underlined in order to make the reader understand that the simplest way to implement and use the MOC is generally not enough in case of more challenging problems.\par
Again, all the three numerical schemes properly capture the exact solutions of the problems. Considering RP1 in Fig.~\ref{fig.RP1}, the less demanding and more general Riemann problem, the Semi-Implicit scheme appears a bit more diffusive along the shock wave than the other two numerical methods. This is due to the parameter \(\theta\) that has to be fixed equal to 0.8 (hence tending to a first order scheme) to avoid oscillations after the rarefaction and before the shock. In RP2 (Fig.~\ref{fig.RP2}) both the DOT and the SI scheme present oscillations in proximity of the contact discontinuity, with the SI having the same flaw also immediately before the shock wave; while the MOC performs in the best way, even if adding diffusion. It is worth to mention that it could be possible to solve the oscillation problem in the Explicit Scheme recurring to a non-linear path, for which the reader can refer to \cite{caleffi2017}. With RP3 the Method of Characteristics demonstrates to have some weak points. In this case, indeed, the scheme is not able to capture the correct evolution of the rarefaction and especially of the shock wave, clearly visible in the velocity plot of Fig.~\ref{fig.RP3}.\par

\begin{figure}[t!]
\begin{subfigure}[b]{0.45\textwidth}
\centering
\includegraphics[width=1\linewidth]{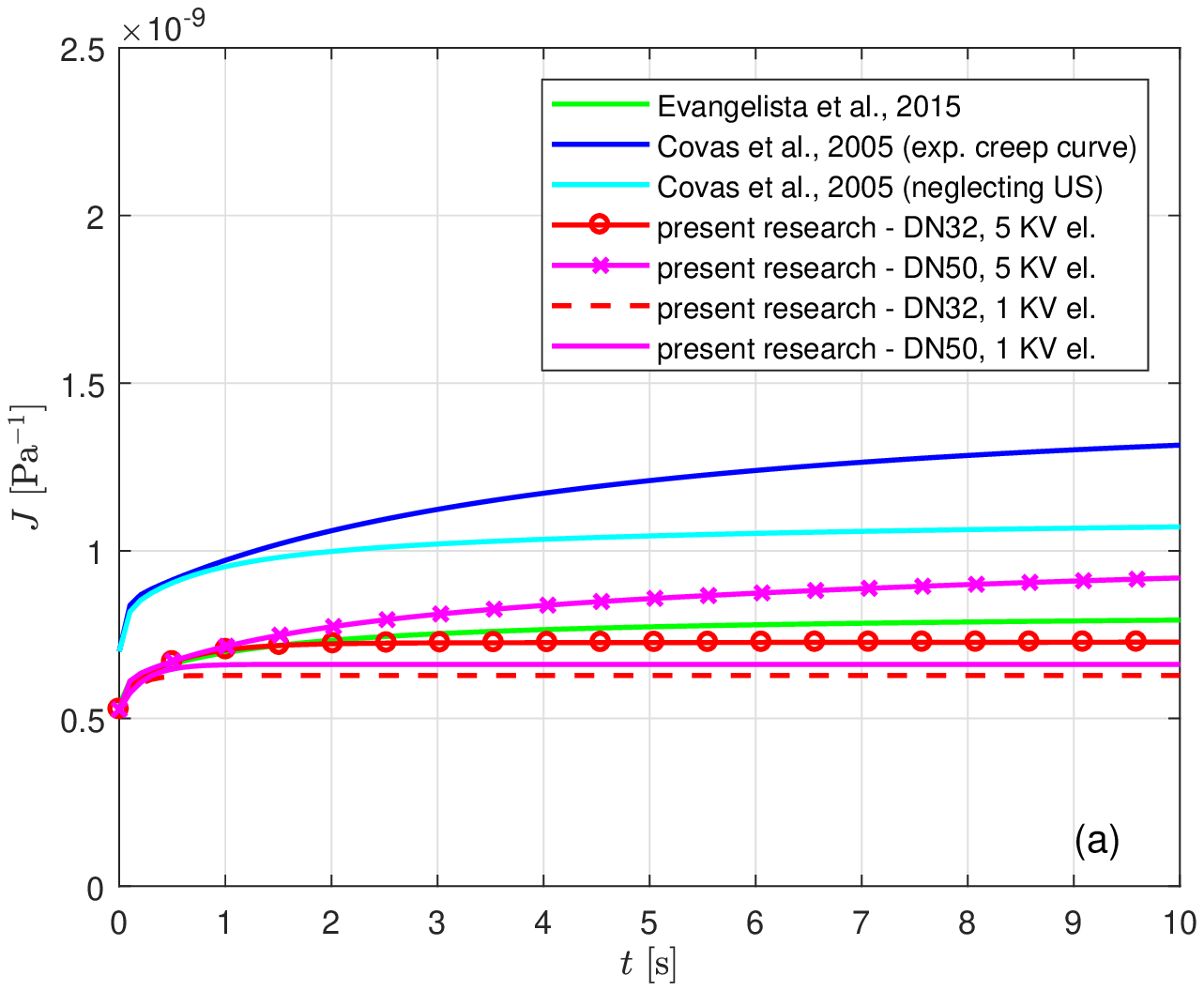}
\label{fig.creep_functions_NOUS}
\end{subfigure}
\hfill
\begin{subfigure}[b]{0.45\textwidth}
\centering
\includegraphics[width=1\linewidth]{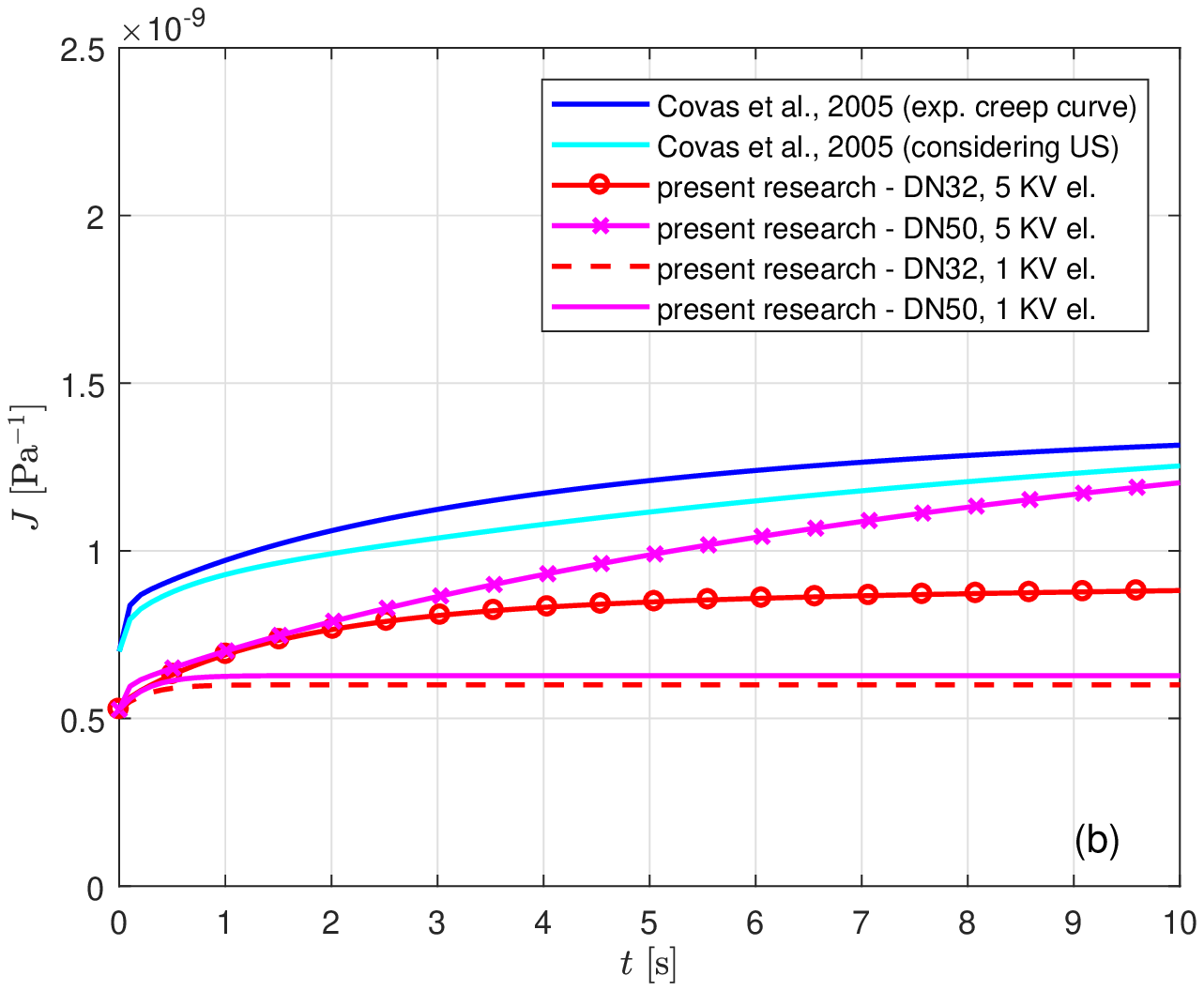}
\label{fig.creep_functions_US}
\end{subfigure}
\caption{Calibrated visco-elastic creep functions of this work, concerning a 3-parameter and a multi-parameter model, compared against previous works' creep functions for HDPE pipes, (a) neglecting unsteady friction and (b) considering unsteady friction.}
\label{fig.creep_functions}
\end{figure}

\begin{table}[b!]
\centering
\begin{tabular}{l P{1.5cm} P{1.5cm} P{1.5cm} P{1.5cm} P{1.5cm} P{1.5cm} P{1.5cm} P{1.5cm}}
\hline
	Test &DN [mm] &$D$ [mm] & $e$ [mm] & $L$ [m] & $Q_0$ [l/s] & $\overline{c}$ [m/s] & $f$ [-]\\
\hline
	\textbf{WH1}  & 50 & 44.0 & 3.0 & 203.3 & 2.00 & 350 & 0.02105 \\
	\textbf{WH2}  & 32 & 23.2 & 4.4 & 101.9 & 0.25 & 500 & 0.03006 \\
\hline
\end{tabular}
\caption{Data of the water hammer test WH1 and WH2.}
\label{tab.WH}
\end{table}\par

\begin{table}[b!]
\centering
\begin{tabular}{l l P{2cm} P{2cm} P{2cm} P{2cm}}
\hline
	Parameter & &\textbf{WH1 - QS} &\textbf{WH1 - US} &\textbf{WH2 - QS} &\textbf{WH2 - US}\\
\hline
	\(E_0\) \quad &[GPa] & 1.90 & 1.90 & 1.90 & 1.90\\
	\(E_{\infty}\) \quad &[GPa] & 1.51 & 1.59 & 1.59 & 1.67\\
	\(\eta\) \quad &[GPa/s] & 0.085 & 0.080 & 0.043 & 0.060 \\
\hline
\end{tabular}
\caption{Visco-elastic parameters calibrated for water hammer test WH1 and WH2 solved with the 3-parameter model in case of a quasi-steady friction model (QS) or considering the unsteady friction losses (US).}
\label{tab.3par}
\end{table}\par

\begin{table}[b!]
\centering
\begin{subtable}{0.7\textwidth}
\centering
\begin{tabular}{l l P{1.5cm} P{1.5cm} P{1.5cm} P{1.5cm} P{1.5cm}}
\hline
	\textbf{WH1 - QS}\\
\hline
	 Parameter & &k = 1 &k = 2 &k = 3 &k = 4 &k = 5\\
\hline
	\(J_k\) &[10$^{-11}$ Pa] & 8.14 & 1.55 & 14.53 & 0.0016 & 23.85 \\
	\(\tau_{rk}\) &[s] & 0.05 & 0.50 & 1.50 & 5.00 & 10.00 \\
\hline
\hline
	\textbf{WH1 - US}\\
\hline
	 Parameter & &k = 1 &k = 2 &k = 3 &k = 4 &k = 5\\
\hline
	\(J_k\) &[10$^{-11}$ Pa] & 6.57 & 0.45 & 3.98 & 0.026 & 89.62 \\
	\(\tau_{rk}\) &[s] & 0.05 & 0.50 & 1.50 & 5.00 & 10.00 \\
\hline
\end{tabular}
\label{tab.multiparDN50}
\end{subtable}
\newline
\vspace*{0.4 cm}
\newline
\begin{subtable}{0.8\textwidth}
\centering
\begin{tabular}{l l P{1.5cm} P{1.5cm} P{1.5cm} P{1.5cm} P{1.5cm}}
\hline
	\textbf{WH2 - QS}\\
\hline
	Parameter & &k = 1 &k = 2 &k = 3 &k = 4 &k = 5\\
\hline
	\(J_k\) &[10$^{-11}$ Pa] & 4.40 & 15.41 & 0.013 & 0.021 & 0.43 \\
	\(\tau_{rk}\) &[s] & 0.05 & 0.50 & 1.50 & 5.00 & 10.00 \\
\hline
\hline
	\textbf{WH2 - US}\\
\hline
	Parameter & &k = 1 &k = 2 &k = 3 &k = 4 &k = 5\\
\hline
	\(J_k\) &[10$^{-11}$ Pa] & 1.91 & 1.00 & 25.64 & 0.93 & 9.78 \\
	\(\tau_{rk}\) &[s] & 0.05 & 0.50 & 1.50 & 5.00 & 10.00 \\
\hline
\end{tabular}
\label{tab.multiparDN32}
\end{subtable}
\caption{Visco-elastic parameters calibrated for water hammer test WH1 and WH2 solved with the multi-parameter model with 5 Kelvin-Voigt elements in case of a quasi-steady friction model (QS) or considering the unsteady friction losses (US), with $E_0=1.90$ GPa.}
\label{tab.multipar}
\end{table}\par

\subsection{Water hammer problems}
\label{S:5.1}

\begin{figure}[t!]
\begin{subfigure}{0.45\textwidth}
\centering
\includegraphics[width=1\linewidth]{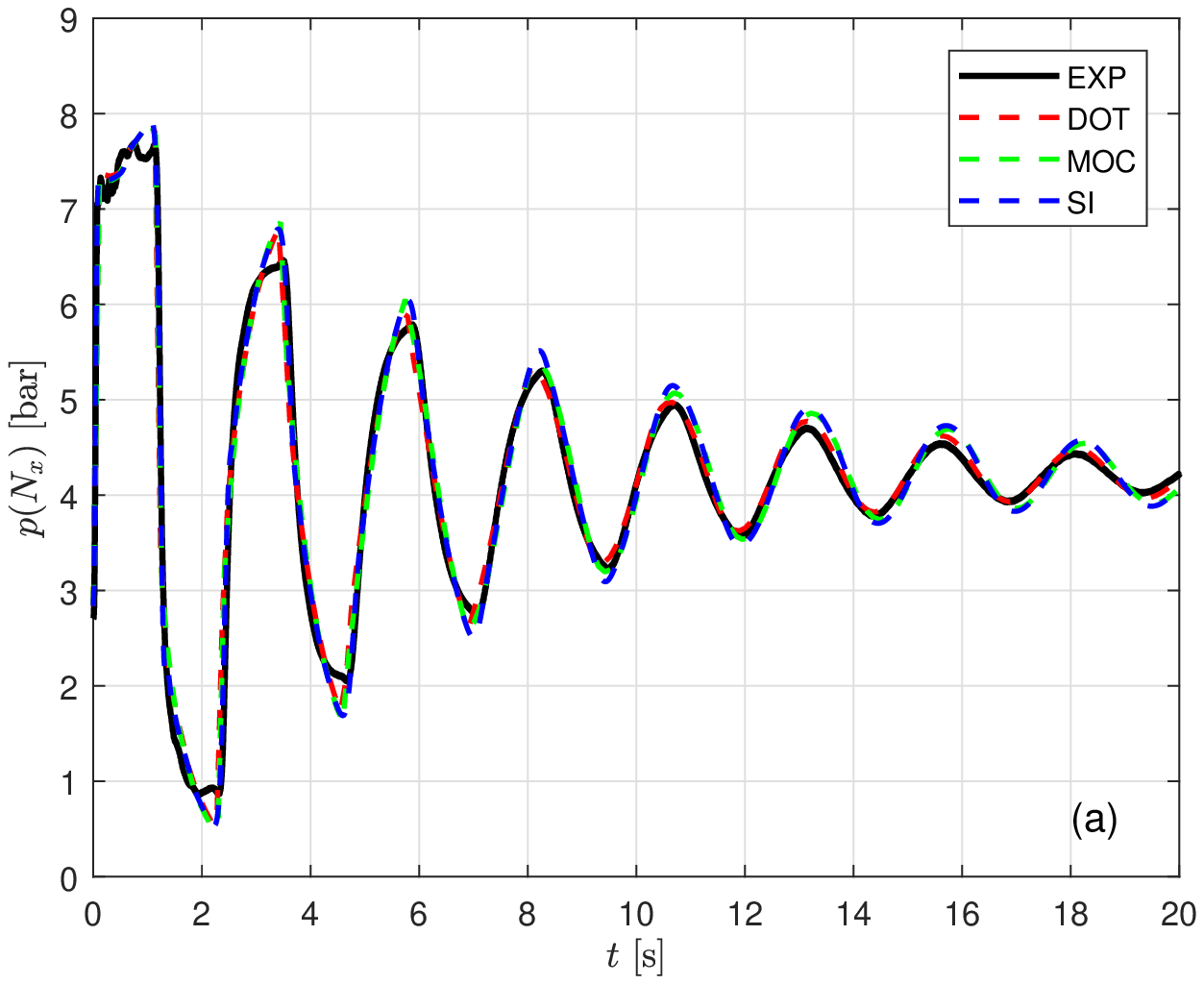}
\label{fig.DN50_3par_steady_out}
\end{subfigure}
\hspace{7mm}
\begin{subfigure}{0.45\textwidth}
\centering
\includegraphics[width=1\linewidth]{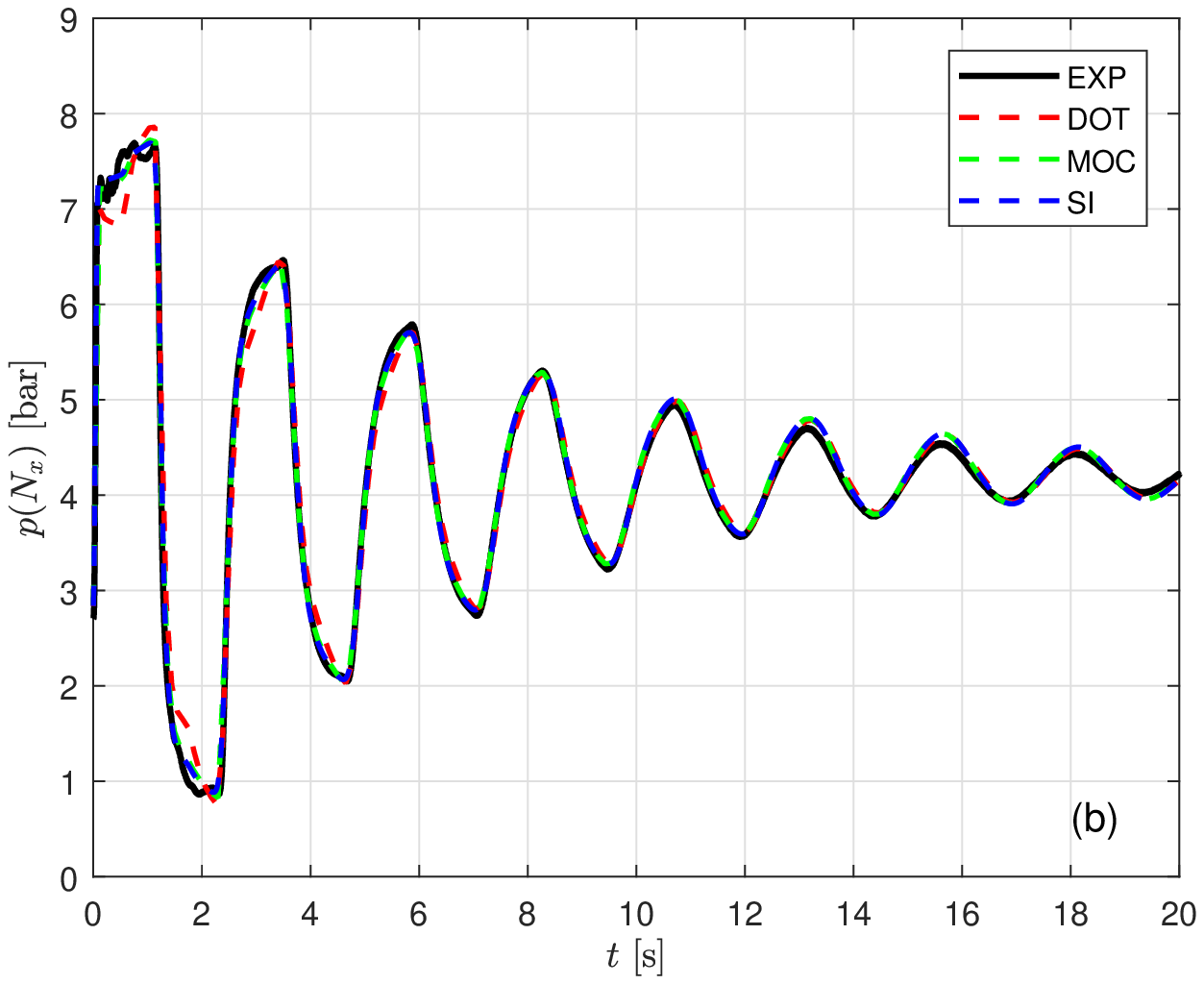}
\label{fig.DN50_multipar_steady_out}
\end{subfigure}
\hfill
\vspace{6mm}
\begin{subfigure}{0.45\textwidth}
\centering
\includegraphics[width=1\linewidth]{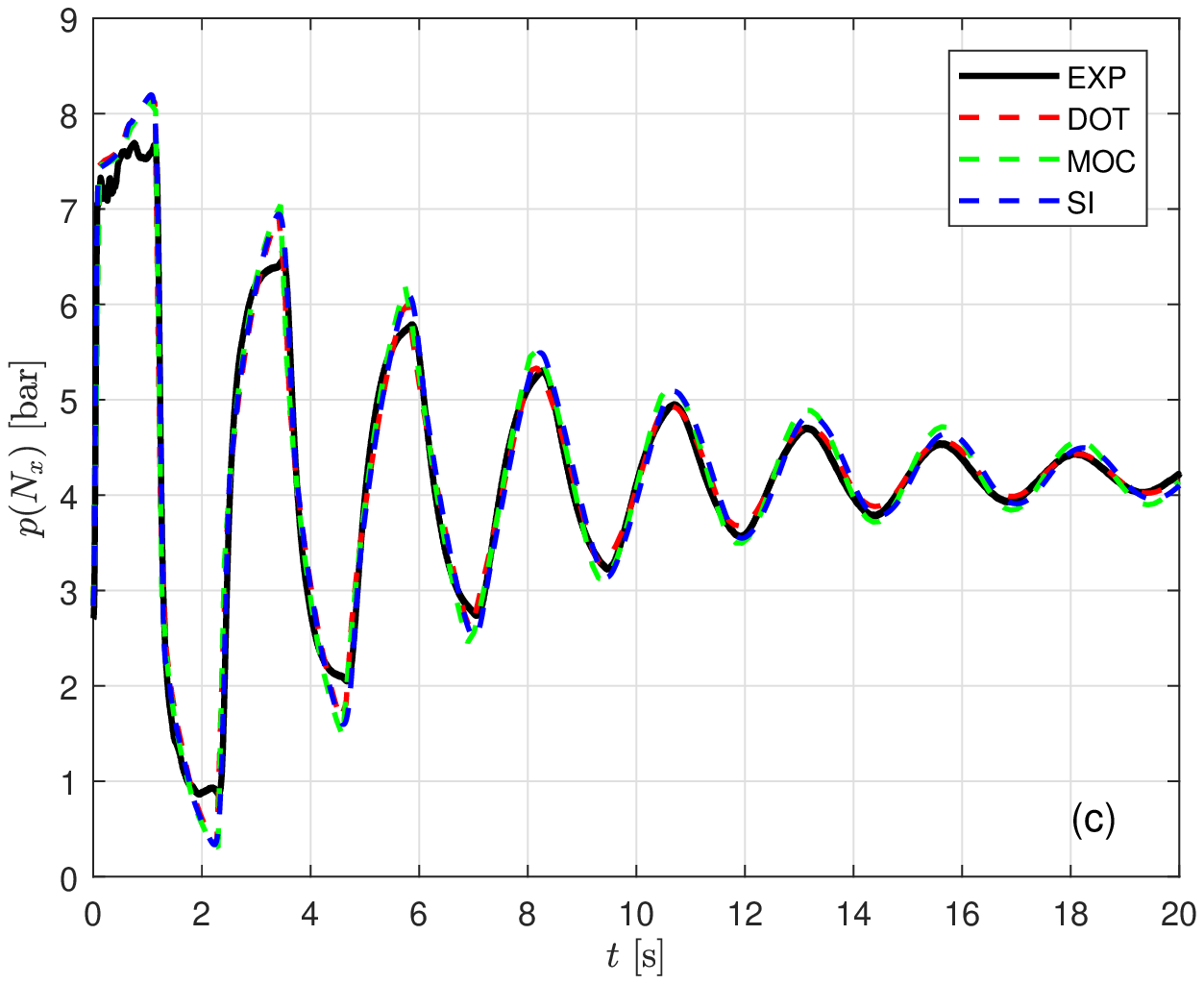}
\label{fig.DN50_3par_unsteady_out}
\end{subfigure}
\hspace{7mm}
\begin{subfigure}{0.45\textwidth}
\centering
\includegraphics[width=1\linewidth]{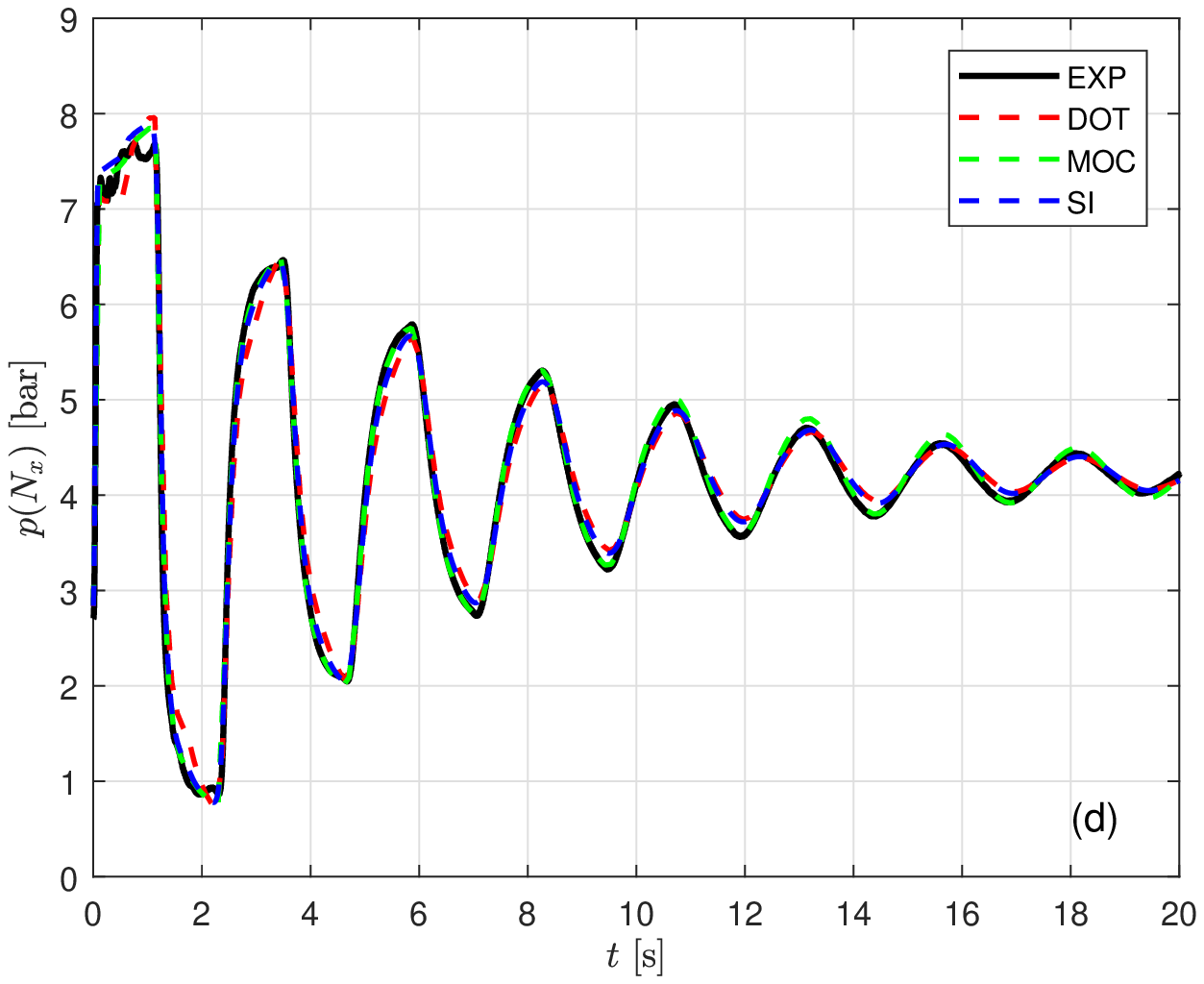}
\label{fig.DN50_multipar_unsteady_out}
\end{subfigure}
\caption{Comparison of the numerical results obtained with MOC, DOT and SI against the experimental solution (EXP) of the water hammer test WH1 with each visco-elastic and friction model configuration: (a) 3-parameter and quasi-steady friction model, (b) multi-parameter and quasi-steady friction model, (c) 3-parameter and unsteady friction model and (d) multi-parameter and unsteady friction model. Pressure \(p(N_x)\) at the downstream end vs time. The reader is advised to refer to the coloured figures of the electronic version of this paper.}
\label{fig.DN50}
\end{figure}

\begin{figure}[t!]
\begin{subfigure}[b]{0.45\textwidth}
\centering
\includegraphics[width=1\linewidth]{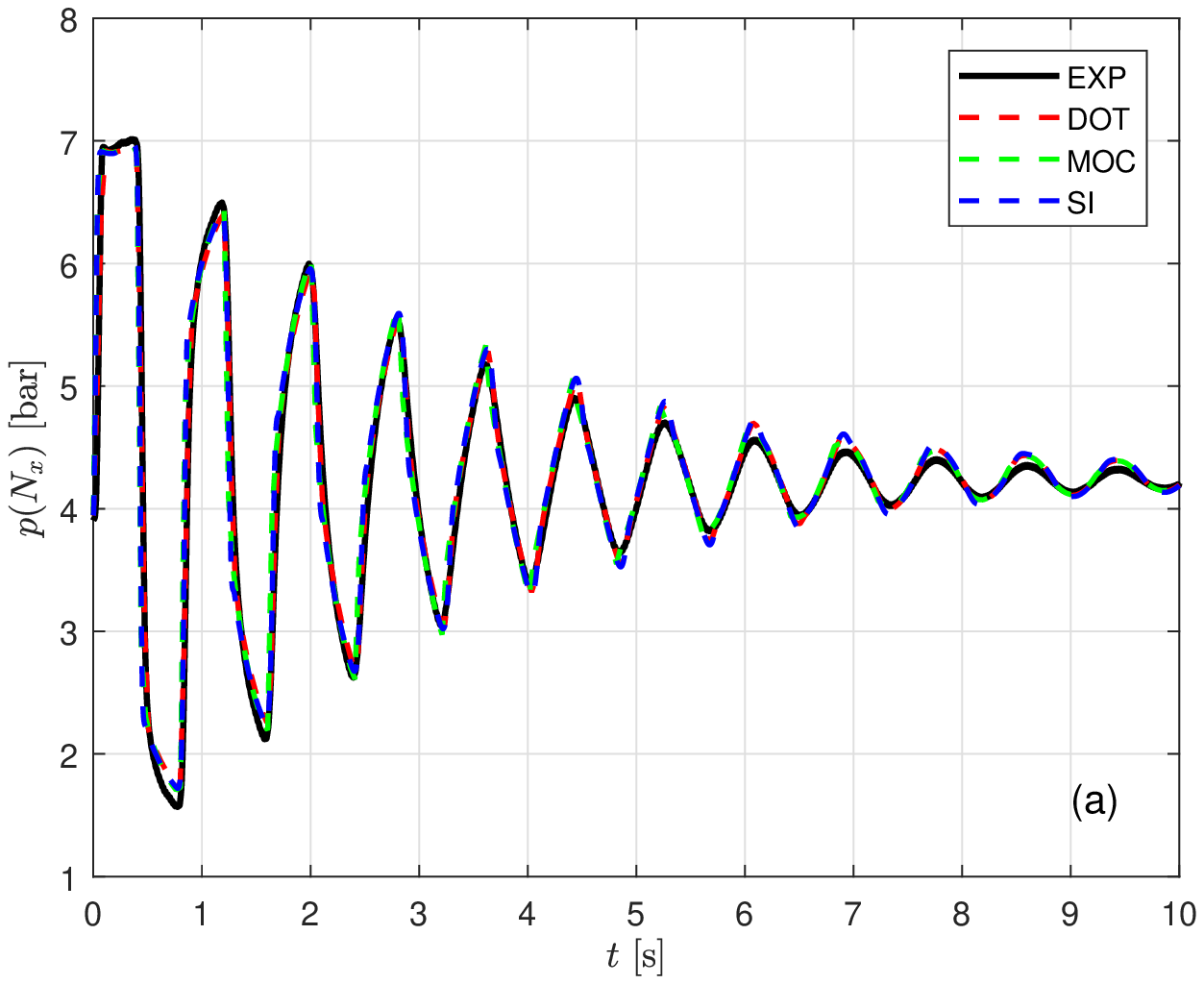}
\label{fig.DN32_3par_steady_out}
\end{subfigure}
\hspace{7mm}
\begin{subfigure}[b]{0.45\textwidth}
\centering
\includegraphics[width=1\linewidth]{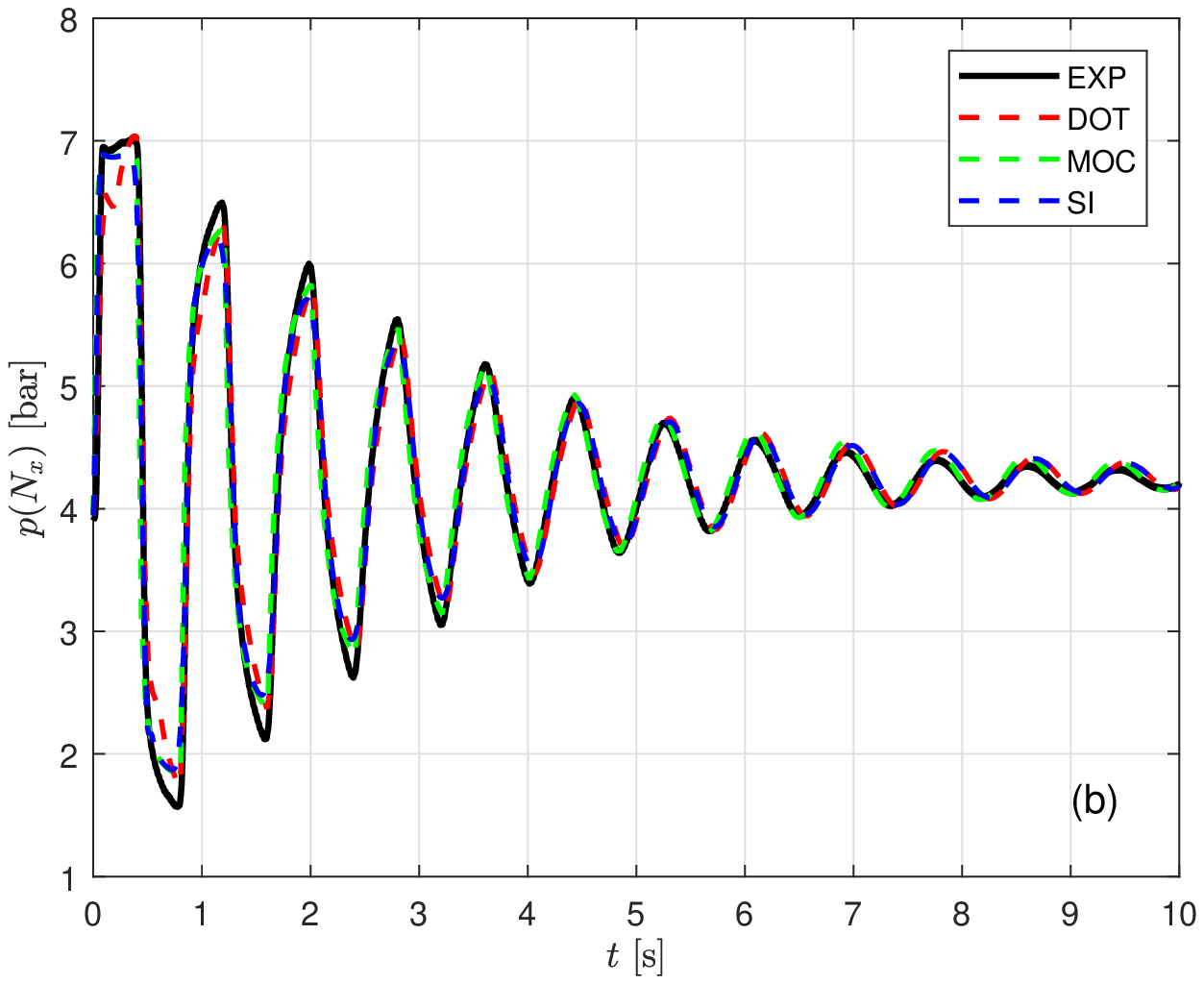}
\label{fig.DN32_multipar_steady_out}
\end{subfigure}
\hfill
\vspace{6mm}
\begin{subfigure}[b]{0.45\textwidth}
\centering
\includegraphics[width=1\linewidth]{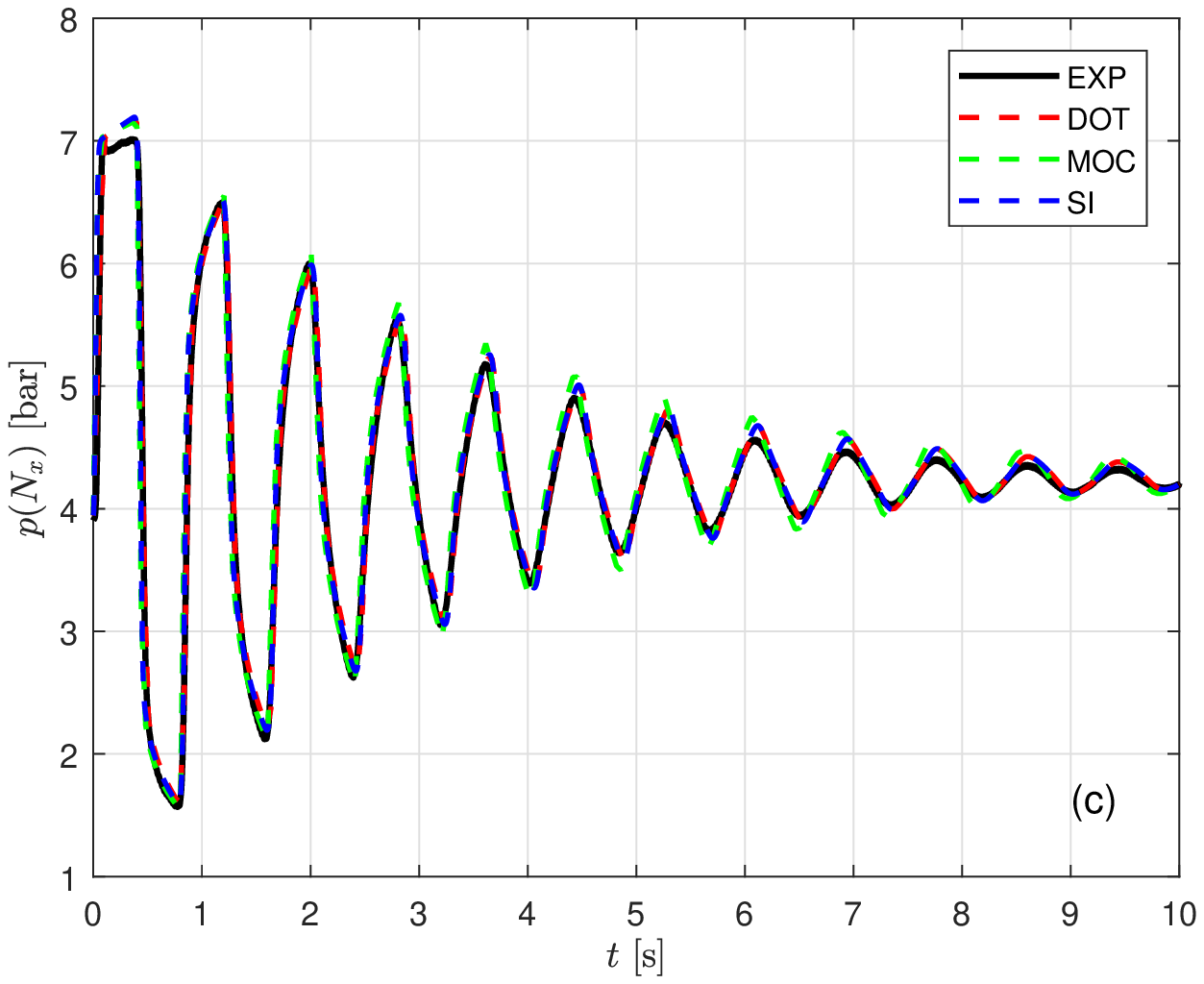}
\label{fig.DN32_3par_unsteady_out}
\end{subfigure}
\hspace{7mm}
\begin{subfigure}[b]{0.45\textwidth}
\centering
\includegraphics[width=1\linewidth]{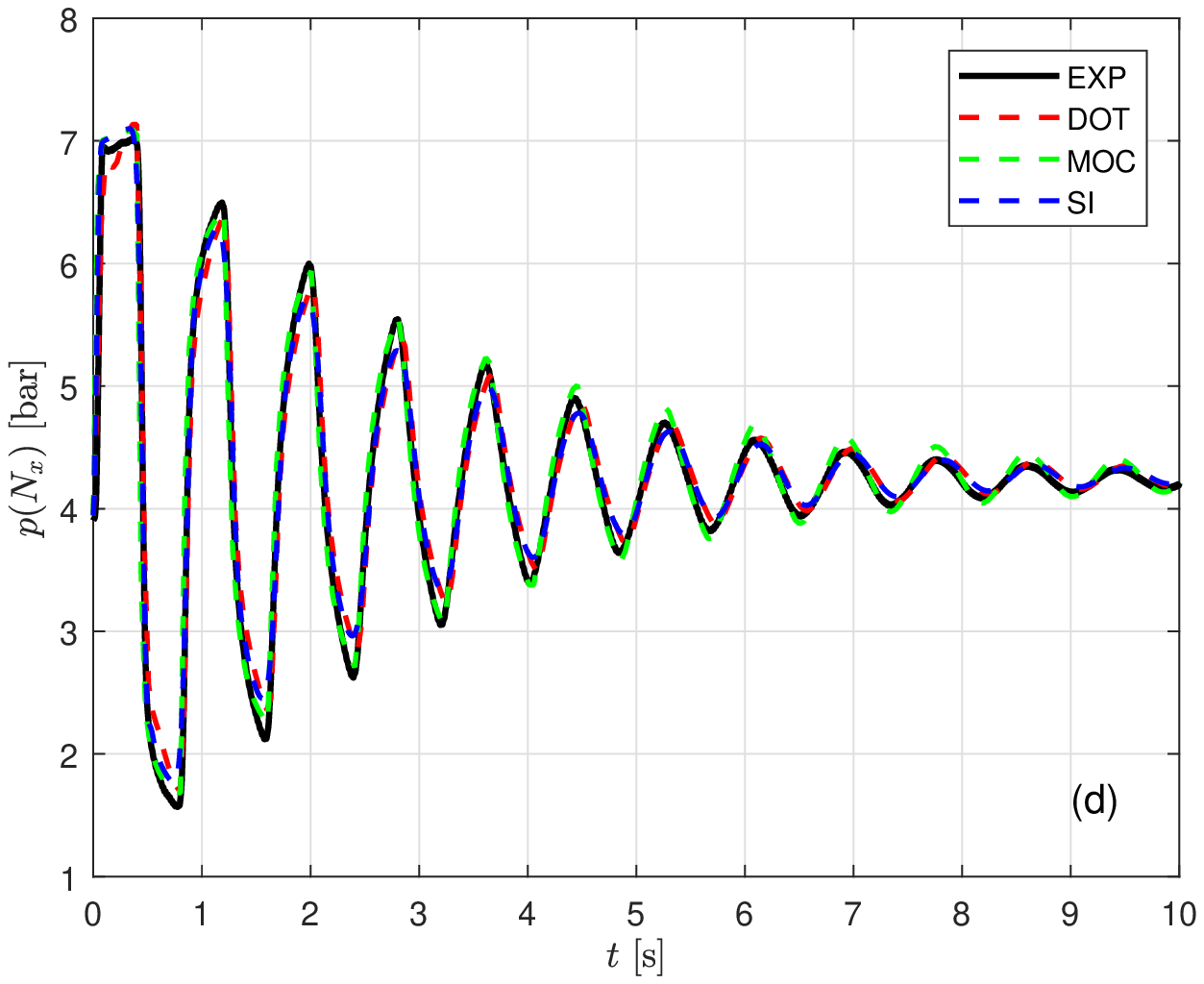}
\label{fig.DN32_multipar_unsteady_out}
\end{subfigure}
\caption{Comparison of the numerical results obtained with MOC, DOT and SI against the experimental solution (EXP) of the water hammer test WH2 with each visco-elastic and friction model configuration: (a) 3-parameter and quasi-steady friction model, (b) multi-parameter and quasi-steady friction model, (c) 3-parameter and unsteady friction model and (d) multi-parameter and unsteady friction model. Pressure \(p(N_x)\) at the downstream end vs time. The reader is advised to refer to the coloured figures of the electronic version of this paper.}
\label{fig.DN32}
\end{figure}

For the water hammer (WH) problems, two HDPE tubes have been chosen, considering the available experimental data. Test WH1 concerns a straight DN50 pipe of length 203.3 m, while test WH2 regards a straight DN32 pipe of 101.9 m. The main features of the systems are listed in Tab.~\ref{tab.WH}. The average wave speeds \(\overline{c}\) were given  from the laboratory experiments, estimated as mean values of those obtained as ratio between four times the total length of the pipe and the time elapsed between two pressure peaks. Pipelines were fixed to the ground by means of metal clamps along the entire length, to avoid any axial movement of the pipes. For both cases, to experimentally generate a transient test, a fast and complete closure of the downstream ball valve was done, with a controlled closure time fixed at 0.1 s (set at the outlet boundary condition). The discharge of the flow was provided upstream from a pressurized tank, whose pressure was measured at each time step and used as inlet boundary condition.\par
To solve these problems considering the correct fluid-structure interaction between water and tube wall, both the 3-parameter and the multi-parameter visco-elastic constitutive models were tested for all the numerical schemes. The visco-elastic parameters (calibrated as explained in section \ref{S:4}) are listed in Tabs.~ \ref{tab.3par}~and~\ref{tab.multipar} for each case and for both the friction models taken into account, the simple quasi-steady (QS) and the unsteady (US) one (using the ODE Model). In Fig.~\ref{fig.creep_functions} it is possible to observe the trend of the calibrated creep functions adopted for this work compared against those used by Evangelista et al. \cite{evangelista2015}, neglecting the unsteady friction effects with a 5 KV elements model, and Covas et al. \cite{covas2004,covas2005}, neglecting the unsteady friction effects or considering them with a 5 KV elements model or using the creep data experimentally determined  in mechanical tests. It has to be mentioned that Covas et al. \cite{covas2005} creep function is referred to a PE pipe and an average pressure wave speed of 395 m/s, corresponding to an instantaneous Young modulus \(E_0 = 1.43\) GPa. It can be noticed that the calibrated creep functions are really comparable to the one proposed by Evangelista et al. \cite{evangelista2015} neglecting the unsteady friction effects. On the other side, considering the unsteady losses, the trend of the curves is similar to those calibrated by Covas et al. \cite{covas2004,covas2005}. In this case it is also visible an increment of the parameters \(J_k\), with respect to the calibration made neglecting the unsteady friction, that confirms a reduction of the elastic modulus \(E_k\) due to the account of the unsteadiness as part of the dampening. The difference between the creep function valid for the DN32 and DN50 conduct can again be attributed to the different facilities and conditions of the two tests: being the visco-elastic models adopted for this work always considerably affected by these aspects, it is generally not possible to fix a unique set of parameters universally valid for a specific material. Finally, if we compare the curves obtained with 1 KV element (3-parameter model) with those with 5 KV elements (multi-parameter model) it can be clearly noticed that adding Kelvin-Voigt elements it is possible to obtain a behaviour of the creep functions that is substantially not constant for higher times. \par
Considering that the parameter $\P$ as defined in \eqref{eq:P} is largely bigger than 1 for the systems analysed in this paper (respectively, $\P = 5.5$ in WH1 and $\P = 12.7$ in WH2), it is initially adopted a quasi-steady friction model inside all the numerical schemes. Nevertheless, we also wanted to test the effect of the unsteady friction model introduced in section \ref{S.S:2.1} (with respect to the steady one) using it inside the schemes.\par
Comparisons between numerical and experimental pressure values in the immediate proximity of the closing valve are shown in Fig.~\ref{fig.DN50} for test WH1 and in Fig.~\ref{fig.DN32} for test WH2 with each visco-elastic model and friction configuration. For all the simulations the number of cells is maintained equal to 50 and with the Semi-Implicit scheme \(\theta = 0.55\) and \(\Delta t_{max} = 0.01\)~s, except for WH2 with the 3-parameter visco-elastic model, for which a smaller \(\Delta t\) was necessary to obtain an accurate result; hence in this case \(\Delta t_{max} = 0.001\)~s. This behaviour can be explained in terms of wave speed: in WH2, indeed, the wave speed is higher than in WH1, meaning that a higher resolution in terms of time steps is necessary if the visco-elastic model adopted is the simplest one. \par
In general it can be noticed that the three numerical methods reproduce similar results in both the test cases.
The first clear observation is related to the contribution of the unsteady friction model, that appears to be negligible, as supposed looking at the parameter $\P$ related to the experiments \cite{ghidaoui2002,duan2010}. This result underlines once more what established by Ghidaoui et al. in \cite{ghidaoui2002}: the unsteady friction term assumes relevance only when the wave has to travel from one end of the pipe to another less than once in order to have the pre-existing turbulent characteristics, throughout the whole cross section of the pipe, influenced by the wall shear pulse.\\
The second remark concerns the visco-elastic models. In our simulations it is visible that the increment of visco-elastic parameters, from 3 to 11, does not yield to a consistent improvement of the results, weighting, on the other hand, in terms of computational costs and adding difficulties to the calibration procedure.\par

\begin{figure}[t!]
\begin{subfigure}{0.45\textwidth}
\centering
\includegraphics[width=1\linewidth]{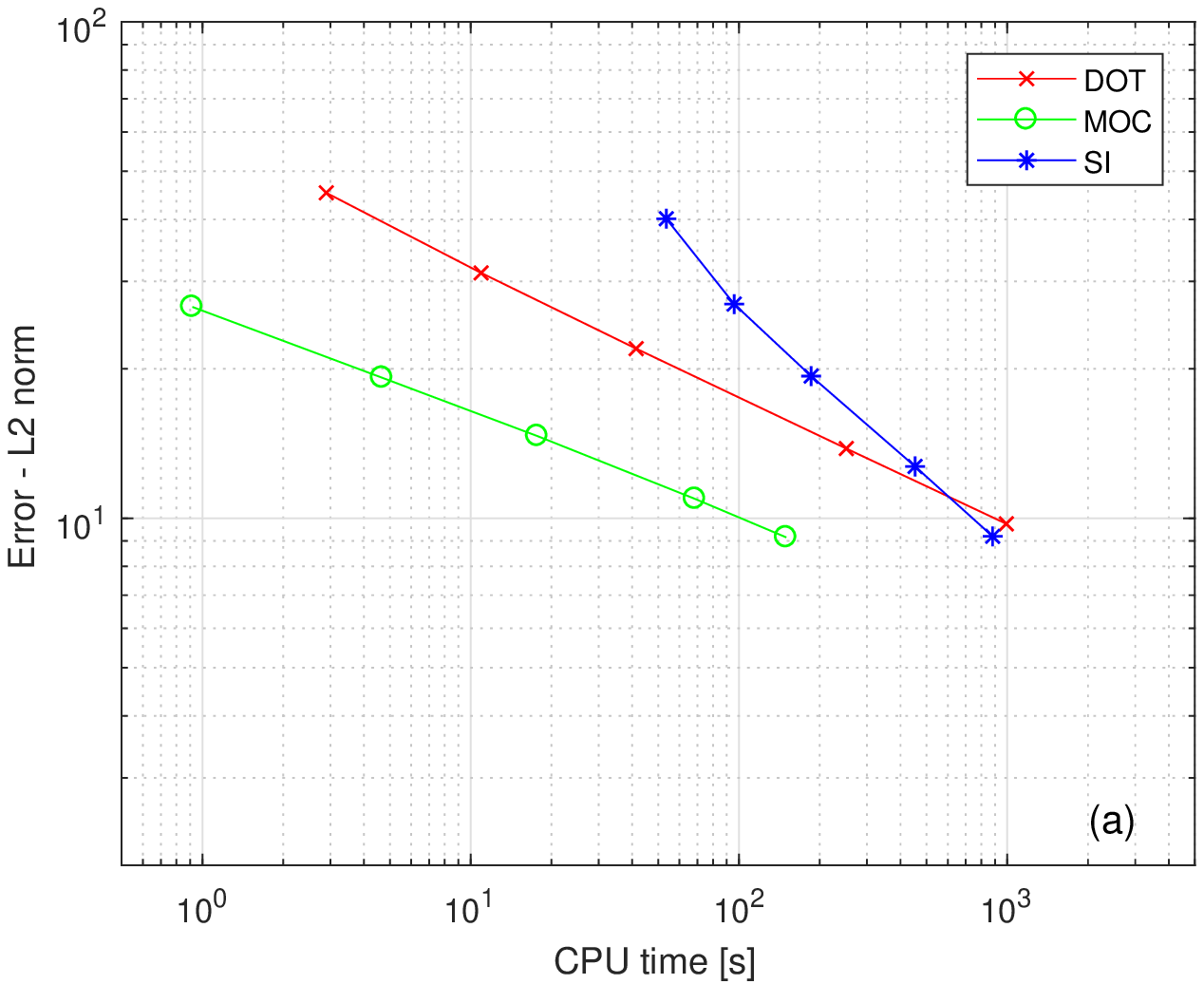}
\label{fig.Efficiency_DN50_3par}
\end{subfigure}
\hfill
\begin{subfigure}{0.45\textwidth}
\centering
\includegraphics[width=1\linewidth]{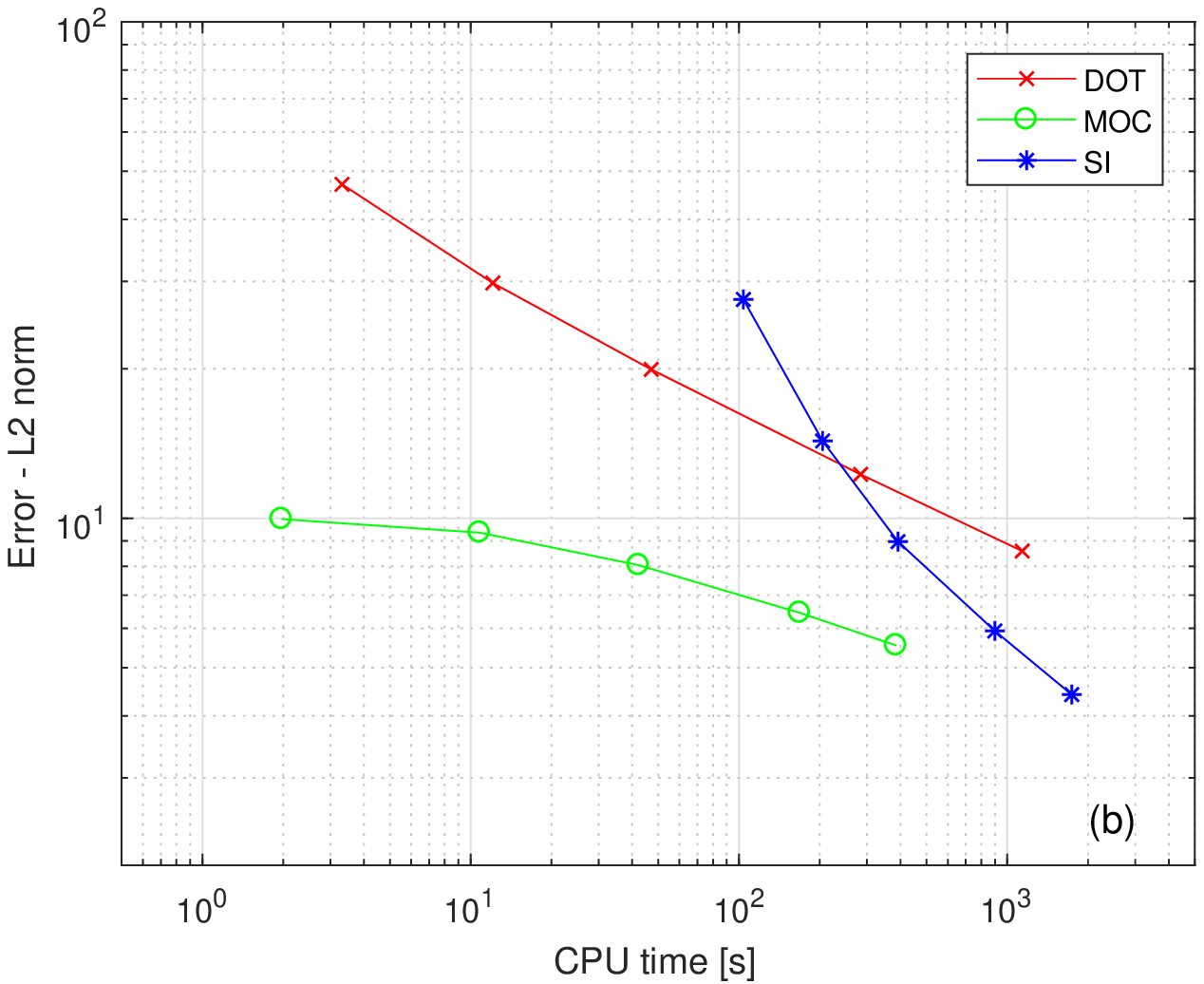}
\label{fig.Efficiency_DN50_multipar}
\end{subfigure}
\caption{Results of the efficiency analysis for the test WH1 with the (a) 3-parameter and (b) multi-parameter visco-elastic model, neglecting unsteady friction; trend of the L2 norm error vs computational time.}
\label{fig.Efficiency_DN50}
\end{figure}

\subsection{Efficiency analysis}
For the water hammer test WH1, an efficiency analysis has been executed to evaluate the performance of the different numerical models. Hence, in Fig.~\ref{fig.Efficiency_DN50} there are compared L2 norm errors against the CPU times separately using the 3-parameter and the multi-parameter visco-elastic models in each scheme. Because of the low impact of the unsteady friction with respect to the accuracy of the results (as observed in section \ref{S:5.1}), with these analysis we consider only a quasi-steady friction model.\\
Solutions are computed for five different meshes: \(N_x = 25,50,100,250,500\) for the DOT and the SI schemes and \(N_x = 100,250,500,1000,1500\) for the MOC (augmented because of the higher efficiency of this numerical scheme).\par
Comparing the two graphs in Fig.~\ref{fig.Efficiency_DN50}, it is evident that increasing the number of visco-elastic parameters to characterize the material leads to an inevitable increment of computational cost, which is not balanced by a comparable error decrement. Considering both the visco-elastic models, it can be clearly deduced that the MOC is the most efficient scheme. The Semi-Implicit Method starts to be competitive only when it is necessary to increase the number of cells of the domain, aiming to obtain very small errors. A parallel observation concerns the trend of each curve: while MOC and DOT maintain almost the same slope (typical of a second order scheme), the SI method presents a steeper slope, meaning a higher order of the scheme. Both these particular behaviours of the SI Method are a consequence of the double condition that has to be respected choosing the maximum admissible \(\Delta t\). With these simulations, the CFL condition is always by-passed by the \(\Delta t_{max}\) fixed in order to avoid excessive numerical diffusion. In this way, the real order of the scheme is hidden and even with a limited number of cells the simulation remains slower than it could be without the fixed time step. Finally, the Explicit Method is always less efficient than the MOC and more efficient than the Semi-Implicit one only when there are fewer cells discretising the domain.\par

\section{Conclusion}
\label{S:6}
The aim of this work is to analyse and compare accuracy, robustness and efficiency of three different numerical schemes, such as Method of Characteristics, Explicit Path-Conservative DOT solver and Staggered Semi-Implicit Finite Volume Method, applied for the resolution of hydraulic transients in flexible polymer tubes. The results show a good agreement with the experimental data for all the numerical methods, whether a Standard Linear Solid Model or a generalized Kelvin-Voigt chain is chosen for the characterization of the visco-elastic mechanical behaviour of the HDPE tube wall. This aspect would tend towards the adoption of less complex 3-parameter models, yet able to adequately capture the correct behaviour of the material and  ensuring in the meantime the minimum computational cost. The same applies concerning the friction term, for which it has been confirmed that, in the scenarios investigated in this paper, the unsteady wall-shear stress can be neglected in favour of a quasi-steady friction model. It is worth remembering that the calibration of the model parameters for the visco-elasticity and for the unsteady friction is complicated by the fact that both these aspects manifest themselves in the damping effect of over-pressure and under-pressure waves. Therefore, a precise calibration of the individual coefficients is hard to achieve. Furthermore, we underline that executing these computational analysis a new efficient resolution of the convolution integral of the unsteady wall-shear stress has been tested in turbulent flow conditions and even an original formulation of the generalized Kelvin-Voigt visco-elastic constitutive law for its applicability to DOT and SI schemes. \\
The most efficient numerical model, among those considered in this study, turns out to be the Method of Characteristics, explaining why it has always been preferred for the resolution of hydraulic transients. Only the Semi-Implicit Method becomes competitive with respect to the MOC when it is necessary to have a rich discretisation of the domain, aiming to obtain very small errors. However, the Riemann problem test cases highlight that the MOC is not as robust as DOT and SI solvers:  to obtain adequate solutions considering more complex configuration in the analysis, such as cross-sectional changes, or more flexible materials (i.e.~rubber), it is not possible to apply the MOC in its simplest way neglecting the convective terms (hence hypothesizing straight characteristic lines). Thus, the code needs to be rearranged for the specific request. Moreover, in the event of cavitation, the Method of Characteristics presents difficulties in the correct capture of the discontinuities inherent in the problem. Therefore, taking into account the considerations here presented, while for simple systems we could easily opt for the MOC, when dealing with complex configurations the choice of the numerical scheme becomes more complicated and it requires the evaluation of the critical aspects involved in the specific case and the maximum error admissible for the results.

\section*{Acknowledgments}
The authors are very grateful to Ph.D. Eng. A. Leopardi and Prof. G. de Marinis (University of Cassino and Southern Lazio), for providing the experimental data of the water hammer test problems presented in this work and to Prof. Alvisi (University of Ferrara) for sharing the MatLab implementation of the SCE-UA algorithm.



\bibliographystyle{abbrv}
\bibliography{MyCollection}



%
%

\end{document}